\documentclass[
 floatfix,
 reprint,
 amsmath,amssymb,
 aps,
]{revtex4-2}
\usepackage[utf8]{inputenc}
\usepackage{amsmath}
\usepackage{physics}
\usepackage{xcolor}
\usepackage{graphicx}
\usepackage{hyperref}
\usepackage{import}
\usepackage{svg}
\usepackage{dcolumn}
\usepackage{bm}

\usepackage[toc,page]{appendix}

\begin{document}

\title{Geometric control of tilt transition dynamics in single-clamped thermalized  elastic sheets}

\author{Roberto Abril Valenzuela}
\affiliation{Department of Physics, University of California, Santa Barbara}
\author{Paul Z. Hanakata}
\affiliation{Department of Physics, Harvard University}
\author{Mark J. Bowick}
\affiliation{Kavli Institute for Theoretical Physics, University of California Santa Barbara}

\begin{abstract}
    We study the finite-temperature dynamics of thin elastic sheets in a single-clamped cantilever configuration. This  system is known to exhibit a tilt transition at which the preferred mean plane of the sheet shifts from horizontal to a plane above or below the horizontal. The resultant thermally roughened two-state (up/down) system possesses rich dynamics on multiple time-scales. In the tilted regime a finite energy barrier separates the spontaneously-chosen up state from the inversion-symmetric down state. Molecular dynamics simulations confirm that, over sufficiently long time, such thermalized elastic sheets transition between the two states, residing in each for a finite dwell time. One might expect that temperature is the primary driver for tilt inversion. We find, instead, that the primary control parameter, at fixed tilt order parameter, is the dimensionless and purely geometrical aspect ratio of the clamped width to the total length of the otherwise-free sheet. Using a combination of an effective mean-field theory and Kramers' theory, we derive the transition rate and examine its asymptotic behavior. At length scales beyond a material-dependent thermal length scale, renormalization of the elastic constants qualitatively modifies the temperature response. In particular the transition is suppressed by thermal fluctuations, enhancing the robustness of the tilted state.  We check and supplement these findings with further molecular dynamics simulations for a range of aspect ratios and temperatures.
\end{abstract}

\maketitle

\section{Introduction}
The properties of polymerized or crystalline membranes (thin elastic sheets) at finite temperature have been of interest for quite some time, both theoretically~\cite{frey_dynamics_1991, le_doussal_self-consistent_1992, aronovitz_fluctuations_1988, NelsonBook, fasolino_intrinsic_2007} and experimentally~\cite{neek2014thermal, blees_graphene_2015, turlier2016equilibrium}. The relative energetic preference  of bending modes over stretching modes, when the system-size exceeds the thickness, leads to scale-dependent material properties of considerable relevance to both soft and  hard condensed-matter systems as well as physical biology. At length scales beyond a material-dependent characteristic length scale, thermal fluctuations play a key role, with non-linear couplings between bending and stretching driving a renormalization of the elastic moduli, the most notable case being the strong growth of bending rigidity with increasing system size.

The implications of thermal fluctuations for the bending and elastic moduli of thin sheets was first understood theoretically \cite{NelsonBook,nelson_fluctuations_1987,aronovitz_fluctuations_1988}. For many years after it was thought that the physical systems most likely to exhibit strong thermal effects would come from the world of soft and/or biological matter, such as the spectrin cytoskeleton of red blood cells. The problem though is that soft materials are easily stretchable as well as bendable and so the dominance of bending over stretching is not manifest until very large length scales, typically larger than the actual physical systems. Hard 2D-metamaterials, such as intrinsically atomically-thin graphene,  are in contrast very stiff to stretching at the microscopic scale (quantum mechanical bonds are electron-volt energy scales) but highly bendable because of their ultra-thinness.  Blees et al. \cite{blees_graphene_2015}, for example, found that micron width graphene ribbons have a bending rigidity four orders of magnitude larger than the microscopic bending rigidity found via first principle calculations~\cite{kudin-PRB-microscopic-bending}. Recent work has also highlighted the subtle role played by boundary conditions and the emergence of purely dimensionless geometric parameters as control variables \cite{hanakata2023vibrations}. There are thermalized versions of the zero temperature instabilities present in classical plate theory, such as the so-called classical Euler buckling \cite{landau1986theory,morshedifard_buckling_2021,shankar_thermalized_2021,kosmrlj_response_2016,hanakata_thermal_2021}.

\begin{figure}[ht]
        \centering 
        \def\svgwidth{\columnwidth}
        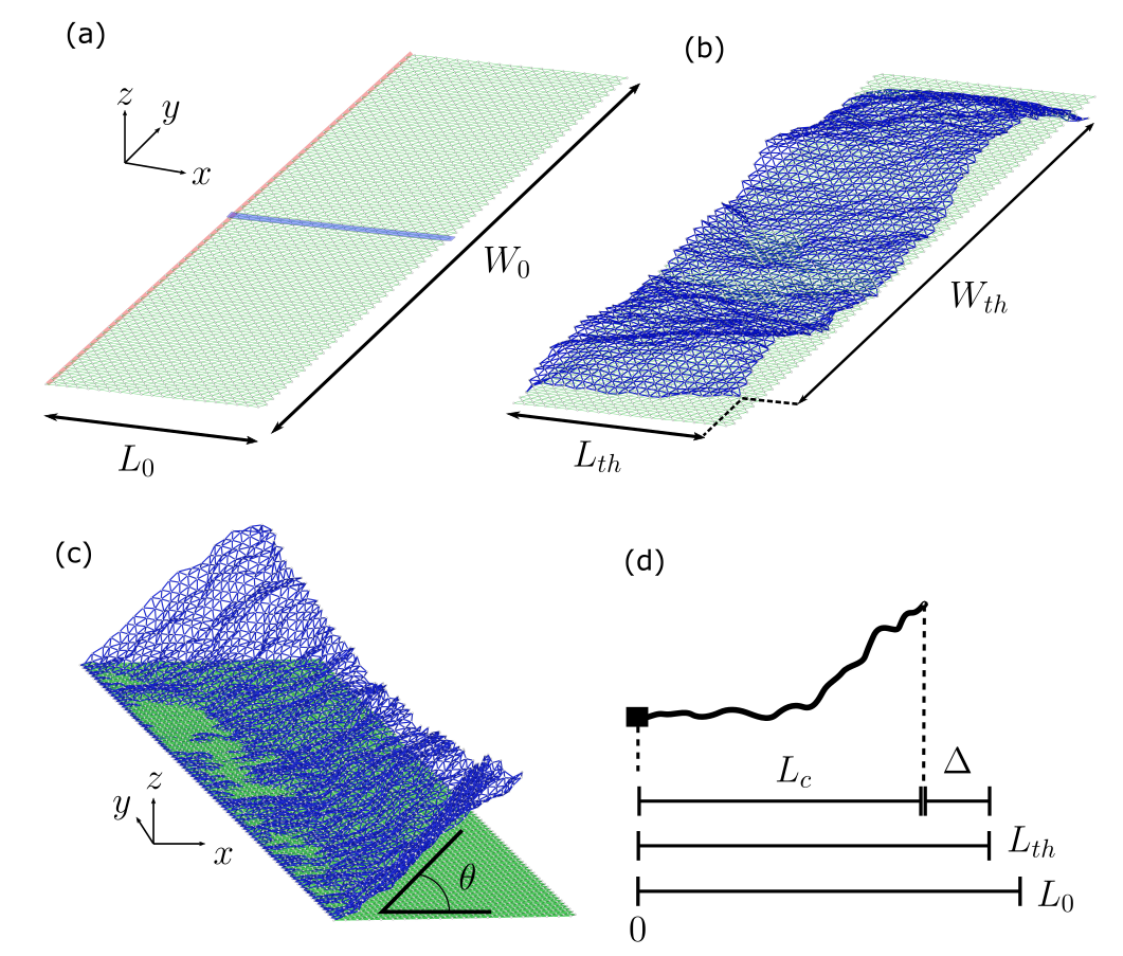
        \caption{(a) $T=0$ triangulated sheet with length $L_0$, width, $W_0$ and aspect ratio $\alpha=W_0/L_0=5$.  Highlighted in red are the fixed vertices corresponding to the clamp at $x=0$. Highlighted in blue and outlined in black are the vertices that constitute the middle slice of the sheet 
        (b) Same sheet, shown in blue, at finite temperature with free boundary conditions: the finite temperature dimensions $L_{th},W_{th}$ are smaller than their zero-temperature counterparts.
        (c) Thermalized sheet with aspect ratio $\alpha=8$ and single-clamped boundary conditions in the tilted state, shown in blue, compared with its $T=0$ configuration, shown in green. Tilt angle $\theta \approx 19^\circ$ with respect to the $z=0$ plane, shown for reference.
        (d) Length scale definitions in the $1D$ representation of the middle slice.}
        \label{sheet}
\end{figure}

A key difference between the thermalized and the zero-temperature system is that the former allows for the existence of internal stresses that facilitate buckling transitions without the need to apply ``extra" strains. This is due to the thermal shrinking of a membrane at finite temperature, leading to a smaller average projected area, $W_{th}\times L_{th}$, as compared to its zero temperature counterpart, $W_0 \times L_0$, at thermal equilibrium (see Fig.~\ref{sheet}b). Thus, clamping at below or above the thermal equilibrium length or width will generate non-zero stresses. 

Earlier work made use of this fact in analyzing thermalized thin sheets in a cantilever configuration.  It was shown, both theoretically and computationally, that clamped boundary conditions along one side of the membrane induce a tilted state in which the mean plane of the sheet is above the horizontal \cite{chen_spontaneous_2022}. This tilted state is more precisely characterized by a nonzero average height at the location of the free edge, $\langle h(x=L,y)\rangle\ne0$, for long time scales. Unlike the classical Euler problem at a fixed length, $L$, however, accessing the tilted state does not depend solely on a critical stress. Instead, it may be induced by varying the dimensionless geometric parameters of the system, in particular the aspect ratio. Spontaneous tilt occurs in thermalized sheets clamped along one edge and only for a window of aspect ratios all above one -- that is the sheet is wider on the clamped edge than it is long.

The basic mechanism driving spontaneous tilt is the following:  with respect to the reference equilibrium state of the free (unclamped) sheet a clamped sheet is under extensile tension concentrated on the clamped edge $(x=0,y)$. This leads to a compressive stress in the orthogonal direction $(x)$ which tends to buckle a sheet with a free end $(x=L)$. This buckling manifests itself as tilt. The degree of the effective clamping stress is directly proportional to the aspect ratio $\alpha=W/L$. Very low $\alpha$ is below the threshold for buckling/tilt and very large $\alpha$ irons out the sheet completely in the horizontal plane (no tilt). Thus tilt occurs for an intermediate range of aspect ratios.

The presence of a tilted phase is supported by a mean field approximation of the critical compression, $\Delta_c$, required to tilt the membrane \cite{chen_spontaneous_2022}. The effective energy is composed of a quadratic stress term and a quartic interaction term, leading to a $\phi^4$-type model with, in the absence of an external field, two degenerate ground states. The two states are separated by a finite energy barrier, allowing thermal fluctuations to drive transitions between the up-and down-tilt states. Indeed, simulations reveal a tilted phase with the expected two-frequency behavior: higher-frequency ($\omega_{well}$) oscillations in a well about a given tilted minimum (say up) and lower-frequency transitions ($\omega_{dwell}$) between the two minima (up to down and vice versa). And so on time scales $t_{well} \propto \omega_{well}^{-1}$ the average height of the free end of the sheet is non-zero and on longer times $\tau_{dwell} \propto \omega_{dwell}^{-1}$ there are up-down transitions that average the height to zero. This is reminiscent of the many examples of two-state systems that exhibit interstate transitions via some form of external energetic kicks, with thermal fluctuations here playing the role of the driving force.
 
Here, we model this tilt transition using mean field theory and invoke Kramers' theory to provide an estimate for the transition rate between a state and its inverted state. We find that for thermalized systems where the system dimensions exceed the characteristic thermal length-scale, temperature plays a surprising role, suppressing transitions as opposed to enabling them. In addition, we find that the transition rate  can be controlled by tuning the dimensions of the system at a fixed temperature. We support this theory by performing molecular dynamics (MD) simulations of the tilt transitions of a triangulated elastic sheet.

\section{Background and model}

\subsection{Elastic Sheets}
Consider an elastic sheet with zero-temperature width $W_0$, length $L_0$ and aspect ratio $\alpha = W_0/L_0$, clamped along one of the two extended edges (for simplicity we choose to clamp the edge at $x=0$). At finite temperature, the free (unclamped) membrane will have equilibrium width $W_{th}$ and length $L_{th}$, both smaller than their $T=0$ counterparts because of the induced thermal topography of the sheet (see Figs.~\ref{sheet}(a) and (b)). Taking this free thermalized state as the appropriate reference state,  we see that clamping one edge is an effective extensional strain focused along the line $(x=0,y)$.

Continuum elasticity theory leads to an elastic free energy of the full thermalized system  of the form \cite{landau1986theory, NelsonBook,chaikin1995principles}
	\begin{eqnarray}
		F[u_{ij}(\mathbf{x}),h(\mathbf{x})] = \int d^2 \mathbf{x}& &\left[\frac{\kappa}{2}(\nabla^2 h)^2 \right.\nonumber \\
		& &\left.+ \mu u_{ij}^2  + \frac{\lambda}{2}u_{kk}^2\right]
        \label{E_cont}
	\end{eqnarray}
where $u_{ij} = (\partial_iu_j+\partial_ju_i+\partial_ih\partial_jh)/2$ is the strain tensor and $u_i$, $h$ are the in-plane and out-of-plane displacements, respectively. The parameter $\kappa$ is the bending rigidity, and $\mu,\lambda$ are Lamé coefficients\cite{landau1986theory}. It is often convenient to quantify the elastic properties of a membrane in terms of the dimensionless Foppl-von Karman number, vK$=Y L^2/\kappa \approx (L/t)^2$, where $t$ is the thickness of the sheet and $Y=4\mu(\mu+\lambda)/(2\mu+\lambda)$ is the 2D Young's modulus. vK measures the ratio of stretching energy to bending energy for a membrane of extended size $L$ and thickness $t$. To get a feel for these values, take for example graphene, with microscopic $\kappa \approx 1.2 \, \text{eV}$ and $Y \approx  20\,  \text{eV Å}^{-2}$: vK is then $\approx  10^{12-13}$ for a sheet of length $L=100$ $\mu$m. Bending deformations are then much less costly, energy-wise,  than in-plane elastic deformations and the high entropy of available bending configurations is a dominant feature of the statistical mechanical response. For the remainder of this paper, we will measure the elastic constants $\kappa$ and $Y$ in units of temperature, $k_BT$. We can then define elastic constants, $\tilde\kappa=\kappa/k_BT$ and $\tilde Y = Y/k_BT$. Temperature will then be measured in terms of scales relative to the scale at which thermal fluctuations become important, $\ell_{th}$. This thermal length scale is usually defined as \cite{nelson_fluctuations_1987,NelsonBook}
    \begin{equation}
        \ell_{th} = \sqrt{\frac{32\pi^3 \kappa_0^2}{3k_BT Y_0}},
        \label{elth}
    \end{equation}
where we denote $\kappa_0,Y_0$ as the bare, microscopic bending rigidity and Young's modulus, respectively. We then set our temperature scales with the dimensionless constant $L/\ell_{th}$, where $L$ is our system size. For reference, at room temperature, the microscopic elastic constants of graphene in units of $k_B T$ are $\tilde\kappa = 48$ and $\tilde Y = 800 \text{Å}^{-2}$. This gives a value for the thermal lengthscale at room temperature of $\ell_{th} \approx 4\text{nm}$. For a lab sample of graphene of size order $L=10\mu\text{m}$, $L/\ell_{th}\sim 10^4$, deep in the thermalized regime. 

\begin{figure}[ht]
        \centering 
        \def\svgwidth{.8\columnwidth}
\begingroup%
  \makeatletter%
  \providecommand\color[2][]{%
    \errmessage{(Inkscape) Color is used for the text in Inkscape, but the package 'color.sty' is not loaded}%
    \renewcommand\color[2][]{}%
  }%
  \providecommand\transparent[1]{%
    \errmessage{(Inkscape) Transparency is used (non-zero) for the text in Inkscape, but the package 'transparent.sty' is not loaded}%
    \renewcommand\transparent[1]{}%
  }%
  \providecommand\rotatebox[2]{#2}%
  \newcommand*\fsize{\dimexpr\f@size pt\relax}%
  \newcommand*\lineheight[1]{\fontsize{\fsize}{#1\fsize}\selectfont}%
  \ifx\svgwidth\undefined%
    \setlength{\unitlength}{296.80611123bp}%
    \ifx\svgscale\undefined%
      \relax%
    \else%
      \setlength{\unitlength}{\unitlength * \real{\svgscale}}%
    \fi%
  \else%
    \setlength{\unitlength}{\svgwidth}%
  \fi%
  \global\let\svgwidth\undefined%
  \global\let\svgscale\undefined%
  \makeatother%
  \begin{picture}(1,0.64355178)%
    \lineheight{1}%
    \setlength\tabcolsep{0pt}%
    \put(0,0){\includegraphics[width=\unitlength,page=1]{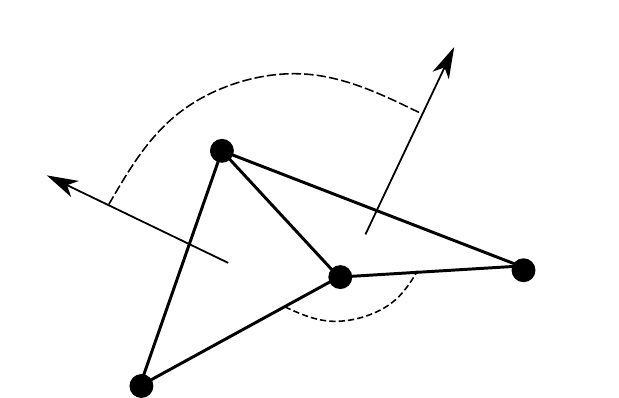}}%
    \put(0.30042621,0.57634064){\makebox(0,0)[lt]{\lineheight{1.25}\smash{\begin{tabular}[t]{l}$\theta_{IJ}$\end{tabular}}}}%
    \put(0.59282491,0.07457016){\makebox(0,0)[lt]{\lineheight{1.25}\smash{\begin{tabular}[t]{l}$\Theta_{dih}$\end{tabular}}}}%
    \put(0.74624396,0.61243932){\makebox(0,0)[lt]{\lineheight{1.25}\smash{\begin{tabular}[t]{l}$\hat{n}_J$\end{tabular}}}}%
    \put(-0.00305992,0.38777059){\makebox(0,0)[lt]{\lineheight{1.25}\smash{\begin{tabular}[t]{l}$\hat{n}_I$\end{tabular}}}}%
  \end{picture}%
\endgroup%

        \caption{Two neighboring plaquettes in the course-grained system forming a dihedral. Simple geometry can be used to replace the angle between the two plaquette normals, $\theta_{IJ}$, by the dihedral angle, $\Theta_{dih}$: $\theta_{IJ}=\pi-\Theta_{dih}$. Using this dihedral angle proves to be more convenient when working with MD packages, such as HOOMD-blue, that have energies associated to dihedral angles readily available.}
        \label{dihedral}
\end{figure}
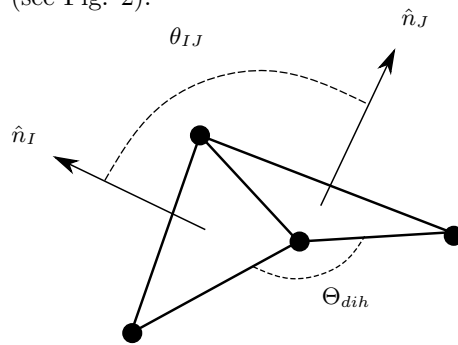

The free energy in Eq.~(\ref{E_cont}) has a discretized energy on a triangular lattice with equilibrium lattice spacing, $a$, of the form ~\cite{seung_defects_1988}
	\begin{equation}
		E = \hat{\kappa}\sum_{\langle I,J\rangle} \left(1-\hat{\mathbf{n}}_I\cdot\hat{\mathbf{n}}_J\right) + \frac{k_{\rm stretch}}{2}\sum_{\langle i,j\rangle} (r_{ij}-a)^{2},
        \label{E_disc}  
	\end{equation}
where the continuum bare bending rigidity $\kappa_0$ and Young's modulus $Y_0$ are related to the discrete bending rigidity $\hat{\kappa}$ and the harmonic spring constant $k_{\rm stretch}$ by $\kappa_0 = \sqrt{3}\hat{\kappa}/2$ and $Y_0 = 2k_{\rm stretch}/\sqrt{3}$. The first term represents the discretized bending energy resulting from normals on adjacent plaquettes (triangular faces) that are not perfectly aligned  and the second term is a harmonic stretching energy between adjacent nodes. The first sum is performed over all nearest neighbor plaquettes, $\langle I,J\rangle$, while the second sum is over all nearest-neighbor vertices $\langle i,j\rangle$ (see Fig. \ref{dihedral}).

\subsection{1D Ribbon Model}

For simplicity, consider a polymer-like approximation to the sheet by taking its midline ($y=0$). Integrating out the  quadratic in-plane modes gives a one-dimensional effective free energy for the height profile $h(x)$:
    \begin{widetext}
	\begin{equation}
		E_{eff}[h(x)] = \frac{\kappa_R W}{2} \int dx\, \left(\dv[2]{h}{x}\right)^2  - \frac{Y_R W}{2L}\Delta  \int dx\left(\dv{h}{x}\right)^2 + \frac{Y_R W}{2L}\left[\frac{1}{2}\int dx\left(\dv{h}{x}\right)^2\right]^2	,
        \label{E_eff}
	\end{equation}
    \end{widetext}
where $\kappa_R$ and $Y_R$ are the renormalized values of $\kappa$ and $Y$, respectively, which scale with system size as
\cite{nelson_fluctuations_1987,NelsonBook,aronovitz_fluctuations_1988}
    \begin{align}
        \kappa_R(\ell) &\sim 
        \begin{cases}
        \kappa_0, \hspace{53pt} \ell<\ell_{th} \\
        \kappa_0 \left(\frac{\ell}{\ell_{th}}\right)^{\eta}, \qquad \ell>\ell_{th} 
        \end{cases}
        \label{kappa_R}
    \end{align}
and 
    \begin{align}
        Y_R(\ell)&\sim
        \begin{cases}
        Y_0, \hspace{64pt} \ell<\ell_{th} \\
        Y_0 \left(\frac{\ell}{\ell_{th}}\right)^{-\eta_u}, \qquad \ell>\ell_{th} .
        \end{cases}
        \label{Y_R}
    \end{align}
where $\ell$ is the length scale over which the sheet is fluctuating. Since long-wavelength fluctuations cannot exceed the smallest macroscopic scale in the problem $\ell\leq L$ for aspect ratio exceeding one. The scaling of the bending rigidity is characterized by the scaling exponent $\eta$ which has been determined  by various analytical methods and numerical simulations to be $\eta\approx 0.8$ \cite{le_doussal_self-consistent_1992, le_doussal_anomalous_2018, NelsonBook}. The exponents $\eta$ and $\eta_u$ are related by rotational invariance: $\eta_u = 2-2\eta$ \cite{aronovitz_fluctuations_1988,bowick_flat_1996}, yielding $\eta_u \approx 0.4$.

The compression of the free end, $\Delta$, is approximately given by \footnote{For a derivation of this result, consult supplemental information of reference \cite{chen_spontaneous_2022})}
    \begin{align}
        \Delta \approx \frac{L_0\alpha \epsilon}{2\sinh^2\left(\frac{\pi\alpha}{4}\right)}\left[\frac{\pi\alpha}{4}\cosh\left(\frac{\pi\alpha}{4}\right)(1+\nu_R)\right.\nonumber\\
        \left.-\sinh\left(\frac{\pi\alpha}{4}\right)(1-\nu_R)\right] ,
        \label{Delta_m}
    \end{align}
where $\nu_R$ is the renormalized Poisson ratio and $ \epsilon\equiv(W_{clamp}-W_{th})/W_{th}$ is the strain at the clamp generated by the thermal shrinking, which is given by \cite{kosmrlj_response_2016}
\begin{equation}
        \epsilon \approx \frac{1}{8\pi\tilde\kappa_0}\left[\frac{1}{\eta}-\frac{1}{\eta}\left(\frac{L_0}{\ell_{th}}\right)^{-\eta}+\ln\left(\frac{\ell_{th}}{a}\right)\right].
        \label{epsi}
    \end{equation}
As in the zero-temperature Euler-buckling problem, one might expect buckling  beyond a threshold compression. Unlike the double-clamp problem, however, the response here is tilt because of an extensive force along the opposite axis produced by the clamped boundary condition {--} stress relaxation then leads to a different buckling mode since there is one free end  \cite{hanakata_thermal_2021}.

\begin{figure}[ht]
        \centering 
        \def\svgwidth{\columnwidth}
        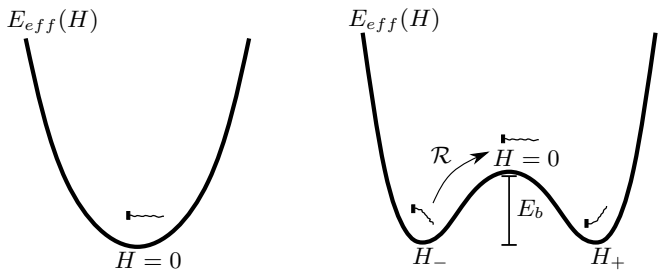
        \caption{Diagramatic depiction of the effective mean-field energy $E_{eff}(H)$ for $\Delta<\Delta_c$ (left) and $\Delta>\Delta_c$ (right). The finite energy barrier,$E_b$, separating the two tilted states allows for transitions from one state to the other with probability $\mathcal{R}$.}
        \label{energy}
\end{figure}

Near the tilt transition: we choose as an ansatz the first buckling mode in the $T=0$ cantilever problem, $h(x) = H \left[1-\cos\left(\frac{\pi x}{2L}\right)\right]$, where $H$ is the height of the free end \cite{timoshenko2009theory}. Upon inserting this  ansatz into the effective energy we obtain a mean field energy
	\begin{equation}
		E_{eff}(H) = a\left(\Delta_c-\Delta\right)H^2 +bH^4
        \label{E_mft}
	\end{equation} 
where $a = \pi^2W Y_R/16 L^2$ and $b= a\pi^2/32L$. This yields a critical compression
    \begin{equation}
        \Delta_c = \frac{\pi^2 }{4L} \frac{\tilde\kappa_R}{\tilde Y_R}.
        \label{Delta_c}
    \end{equation}
In this form, it is easy to see a clear separation between the flat phase ($\Delta<\Delta_c$) and the tilted phase ($\Delta>\Delta_c$).

 \begin{figure*}[t]
    \centering 
    \def\svgwidth{\textwidth}
\begingroup%
  \makeatletter%
  \providecommand\color[2][]{%
    \errmessage{(Inkscape) Color is used for the text in Inkscape, but the package 'color.sty' is not loaded}%
    \renewcommand\color[2][]{}%
  }%
  \providecommand\transparent[1]{%
    \errmessage{(Inkscape) Transparency is used (non-zero) for the text in Inkscape, but the package 'transparent.sty' is not loaded}%
    \renewcommand\transparent[1]{}%
  }%
  \providecommand\rotatebox[2]{#2}%
  \newcommand*\fsize{\dimexpr\f@size pt\relax}%
  \newcommand*\lineheight[1]{\fontsize{\fsize}{#1\fsize}\selectfont}%
  \ifx\svgwidth\undefined%
    \setlength{\unitlength}{594.54719441bp}%
    \ifx\svgscale\undefined%
      \relax%
    \else%
      \setlength{\unitlength}{\unitlength * \real{\svgscale}}%
    \fi%
  \else%
    \setlength{\unitlength}{\svgwidth}%
  \fi%
  \global\let\svgwidth\undefined%
  \global\let\svgscale\undefined%
  \makeatother%
  \begin{picture}(1,0.50105681)%
    \lineheight{1}%
    \setlength\tabcolsep{0pt}%
    \put(0.05412211,0.48572215){\makebox(0,0)[lt]{\lineheight{1.25}\smash{\begin{tabular}[t]{l}(a)\end{tabular}}}}%
    \put(0.604966,0.47706785){\makebox(0,0)[lt]{\lineheight{1.25}\smash{\begin{tabular}[t]{l}(b)\end{tabular}}}}%
    \put(0.61297373,0.24423334){\makebox(0,0)[lt]{\lineheight{1.25}\smash{\begin{tabular}[t]{l}(c)\end{tabular}}}}%
    \put(0,0){\includegraphics[width=\unitlength,page=1]{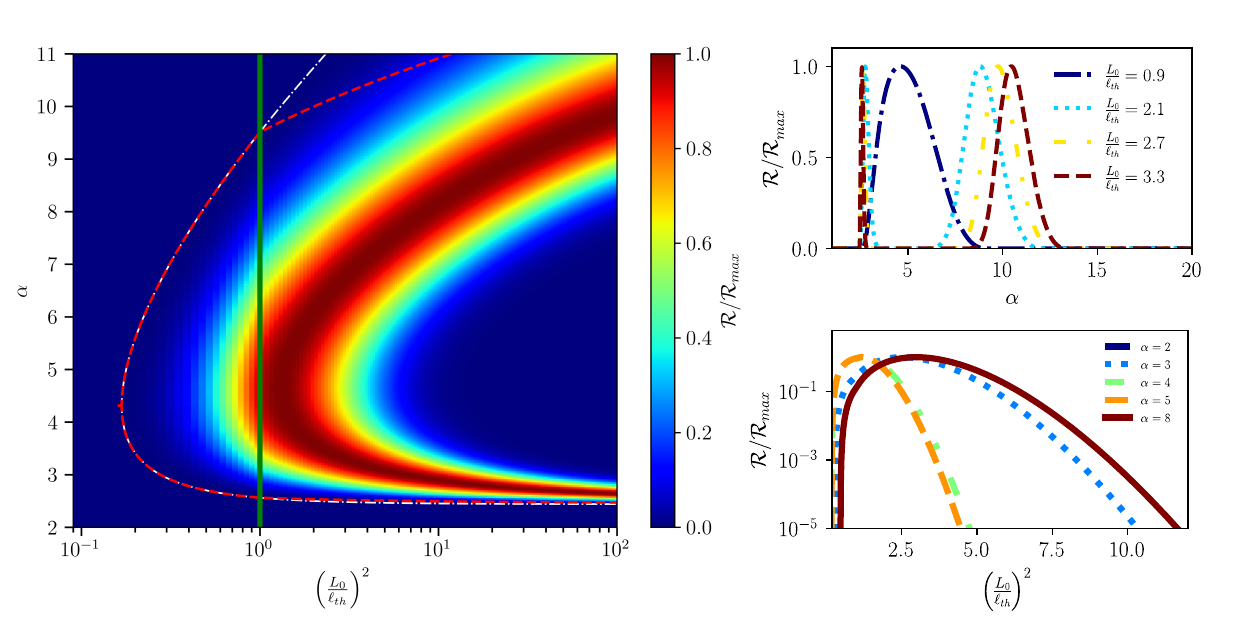}}%
  \end{picture}%
\endgroup%

    \caption{(a) Density plot of the theoretical transition rate as a function of aspect ratio and temperature ($\propto (L/\ell_{th})^2$) with color, as indicated by the color bar (right), representing the estimated transition rate normalized by the maximum value within the range. Mean field theory prediction of the tilted phase boundary is shown as a red dashed line, where we differentiate between bare and renormalized elastic constants beyond the $L_0/\ell_{th}=1$ line (green solid line). The white dash-dotted line shows the phase boundary were the elastic constants to have no scale dependence beyond the thermal length scale. (b-c) 2D slices of the theoretical transition rate as a function of aspect ratio (b) and temperature(c).}
    \label{phaseplot}
\end{figure*}
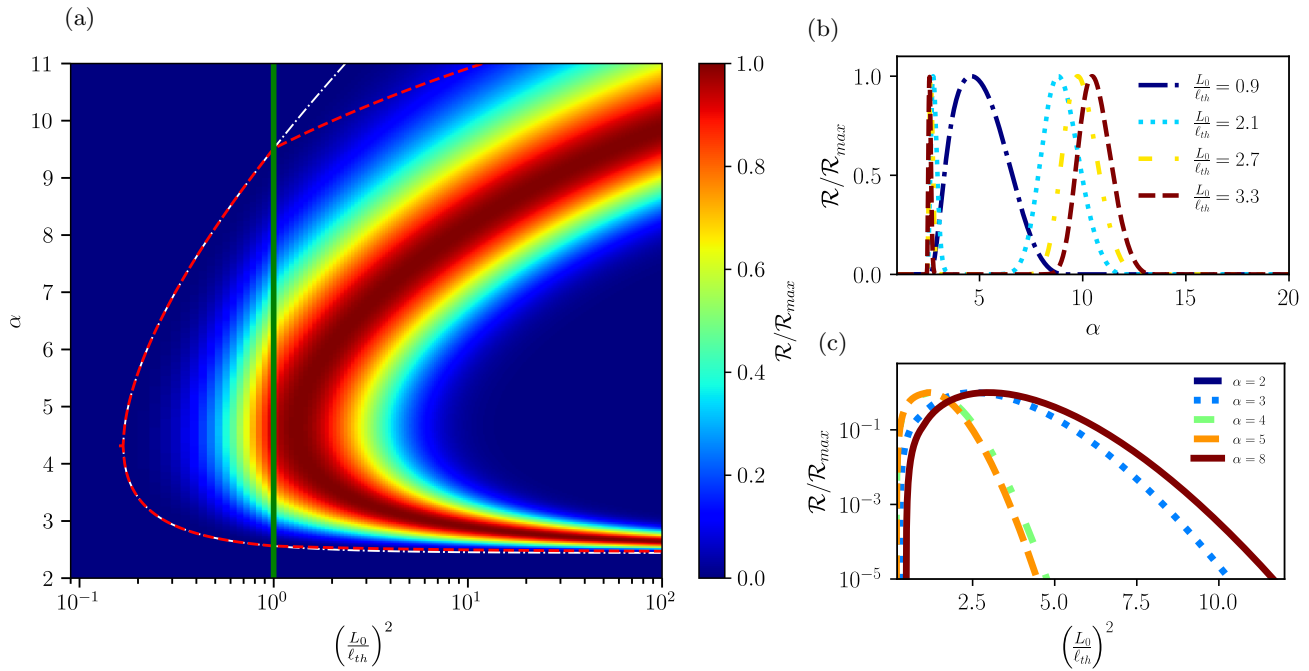

In the tilted phase there are two minima, $E_{\pm}$, separated by an energy barrier $\Delta E_b = |E_{flat}-E_\pm|$, where $E_{flat}=E(H=0)=0$ is the energy of the unstable flat state and $E_{\pm}=E(H_{\pm})$ is the energy of a tilted state (see Fig. \ref{energy}). Once the system is in one of the tilted states, as in any such two-state systems \cite{feynman2011feynman}, there is a non-zero probability of transitioning from one state to the other with maximal transition probability at some resonant value of an external parameter such as temperature or an external driving  frequency ~\cite{RevModPhys_stochastic_resonance}. One might expect the transition rate  at finite temperature to be controlled primarily by thermal fluctuations over the barrier. We show here, however, that it is the dimensionless and purely geometrical aspect ratio that is key determiner.

\subsection{Transition Rates}
Consider a system with an energy landscape given by Eq.~\ref{E_mft} and assume that $\Delta>\Delta_c$ so that we are in the tilted state. We can assign each of the extrema of $E_{eff}$ a characteristic frequency, $\omega_\pm$ and $\omega_B$, which are obtained from the second-order expansion of $E(H)$ at one of the tilted states ($H=H_{\pm}$) and the saddle point ($H=0$), respectively. We can estimate the rate $\mathcal{R}$ of transitioning from one of the tilted wells to the other using Kramers' theory \cite{zwanzig2001nonequilibrium,amir2020thinking}, which predicts 
    \begin{equation}
	   \mathcal{R} \approx R_0 e^{- \Delta E_b/k_BT }
        \label{R_kramers}
    \end{equation}
where the amplitude $R_0$ will take a form that depends on the friction of the system, $\beta= \gamma/m$, which is the ratio of the friction to the mass $m$ of the sheet and has the units of frequency. The system can be underdamped ($\beta\ll\omega_B$) or overdamped ($\beta\gg \omega_B$). Minimizing the free energy (\ref{E_mft}) gives the location of the minima, $H_{\pm}$, and the oscillation frequencies within a minimum,  as well as the height of the barrier. The depth of the energy barrier is given by
    \begin{equation}
         \Delta E_b = \frac{\pi^4W\kappa_R^2}{32 Y L^3}\left(\frac{\Delta-\Delta_c}{\Delta_c}\right)^2
         \label{E_b}
    \end{equation}
All together, this yields a transition rate
    \begin{equation}
		\mathcal{R} \approx 
				R_0\exp\left[-\frac{\pi^4\bar{\Delta}^2 W}{32L^3}\frac{\kappa_R^2}{k_B T Y_R}\right],
        \label{R_pretherm}
    \end{equation}
for relative compression $\bar{\Delta}\equiv (\Delta-\Delta_c)/\Delta_c$, which is positive in the tilted phase.  Note that the energy landscape is symmetric about the flat state (Fig. \ref{energy}), which means {that transition rates are also symmetric with respect to inversion.} We can generalize this by adding a symmetry-breaking, transverse field (such as a gravitational or an electric field) that couples linearly to the height $h(x)$ in Eq~(\ref{E_cont}). This will create an asymmetric potential well, resulting in two distinct transition rates.

We are interested in system sizes sufficiently large that thermal fluctuations are important. This means the length of the sheet satisfies $L\gg\ell_{th}$, where $\ell_{th}$ is the characteristic thermal length scale beyond which the elastic constants become scale-dependent.

{For a thermalized system the elastic moduli $\kappa$ and $Y$ are renormalized by thermal fluctuations, rendering them length-scale dependent \cite{NelsonBook}.} They must then be replaced by their respective renormalized values given by the scalings in Eqs.~\ref{kappa_R} and ~\ref{Y_R}. 

{With these scalings the transition rate simplifies to}
    \begin{equation}
       \mathcal{R} \approx R_0 \exp\left(-\frac{3\pi\bar{\Delta}^2}{512}\alpha\right), 
       \label{R_therm}
    \end{equation}
for $L\gg\ell_{th}$. For a fixed relative compression $\bar{\Delta}$, corresponding to a fixed value of the tilt order parameter or energy barrier, the transition rate is controlled by the aspect ratio $\alpha$ in the exponential, which therefore plays the role of a Boltzmann factor. Effectively geometry is replacing temperature. Temperature enters implicitly in tuning to a fixed relative compression as well as in the amplitude.

As previously mentioned, the prefactor $R_0$ depends on the magnitude of the friction, $\beta$. It takes the form \cite{zwanzig2001nonequilibrium} 
    \begin{equation}
        R_0 \approx \begin{cases}
            \frac{m\beta \Delta E_b}{\pi k_B T}, \qquad \beta\ll \omega_B\\
            \frac{\omega_{\pm}\omega_{B}}{2\pi \beta},\qquad \beta\gg \omega_B,
        \end{cases}
        \label{R0_cases}
    \end{equation}
where we see a turnover from a linear to an inversely proportional dependence in $\beta$. In the thermalized limit, $L\gg \ell_{th}$, we obtain the following prefactors 
    \begin{equation}
        R_0 \approx \begin{cases}
            \frac{3m\beta\bar{\Delta}^2}{512}\alpha, \hfill \beta\ll \omega_B\\
            \frac{\pi^3\bar{\Delta} \kappa_0 }{32\sqrt{2}m\beta L^2}\left(\frac{L}{\ell_{th}}\right)^{\eta}\alpha,\qquad \beta\gg \omega_B,
        \end{cases}
        \label{R0_thermalized}
    \end{equation}
in which we note that the temperature dependence in the overdamped case comes from the renormalization of the bending rigidity as given by Eq.~\ref{kappa_R}. In the underdamped case, we see the same cancellation of temperature that occurs in the Arrhenius factor in Eq.~\ref{R_therm} and we have explicit independence of temperature at constant compression $\Delta$. Note that for both cases, we expect the Arrhenius factor to provide the dominant behavior for significant compression $\bar\Delta$.

Fig. \ref{phaseplot} shows a density plot of the underdamped transition rate as a function of temperature and aspect ratio, normalized by the maximum rate in the displayed region as predicted by Eqs.~\ref{R_pretherm}-~\ref{R0_thermalized}. We include a phase boundary provided by setting the predicted compression, $\Delta$ in Eq.~\ref{Delta_m}, equal to the critical compression, $\Delta_c$, computed from the 1D mean field theory (Eq.~\ref{Delta_c}). In other words, we plot the line $\Delta=\Delta_c$ accounting for renormalization of the elastic constants for $L_0>\ell_{th}$. To the left of this boundary, the system is in the flat phase ($\Delta<\Delta_c$) and the transition rate vanishes. Note the high transition rate localized near the phase boundary. As the temperature is increased deep in the tilted state the transition rate reaches a dynamically stable ``basin" with long dwell times, as indicated by the dark blue region. The $\alpha$-scaling form of Eq.~\ref{R_therm} does not capture the full behavior near the upper branch of the phase boundary as it neglects the higher order contributions coming from $\bar{\Delta}$. Indeed  our mean field theory will break down in the limit of large $\alpha$ where the length of the membrane becomes negligible compared to the width and the 1D midline  model is no longer applicable.

Temperature is a more traditional parameter to tune the dynamics of these types of oscillator systems as higher temperatures often decrease the energy barrier and allows for higher transition probabilities. In this system, however, we see that thermalization gives access to another possible parameter that controls transitions namely, $\alpha$, and with the addition of thermal factors in the prefactor, we see that high temperature instead stabilizes the system in one of the stable tilted states with close to zero transition probability. This can be thought of as temperature decreasing the rest area of the reference state, thereby effectively increasing the clamping strain. Based on the predicted form of the transition rate, we can expect $\mathcal{R}\approx 0$ for large temperatures, where we expect the sheet to be in a fully tilted state. This means that in an experimental setting, low temperatures are needed to access the dynamic tilted state, where $\mathcal{R}\ne0$, and in order to retain a specific constant transition rate, the temperature must remain constant. One can circumvent this by considering the aspect ratio of the sample as a tunable parameter. We can prescribe appropriate geometrical dimensions corresponding to an aspect ratio that leads to the desired transition rate at some constant temperature. Thus, the aspect ratio gives access to a larger sample parameter space in the production of nano-mechanical actuators and may prove to be are more desirable parameter to tune in the manufacturing of such devices.

\section{Molecular Dynamics Simulations}

\subsection{Simulation Setup}

The coarse-grained energy (\ref{E_disc}) can be simulated using the HOOMD-blue python package for molecular dynamics (MD) \cite{anderson_hoomd-blue_2020} for a triangular lattice with $a=1$. The stretching term {is treated} as a harmonic potential between two {nodes} and the bending energy {has a discrete representation} in terms of the dihedral angle formed by neighboring triangular plaquettes, $\Theta_{dih}=\pi-\theta_{IJ}$, where $\theta_{IJ}$ is the angle between the two plaquette normals (see Fig.\ref{dihedral}). Dihedral angles can be readily obtained using HOOMD-blue. A triangular lattice with fixed dimensions is initialized, clamped at one edge, and then integrated in an NVT ensemble for a total of $N=2\times10^7-10^8$ time steps with step size $dt=0.005$ time steps and with energy scale set by $k_B T$. The first half of these time steps is discarded to ensure thermalization. We extract the time series of the out-of-plane height of a single node at the middle of the free edge opposite the clamped edge: $(x,y)=(L,0)$. We generate multiple independent runs ($n= 3$-$5$) for each set of initial parameters to generate error statistics which are computed using the jackknife procedure \cite{amit2005field}.

To test the predicted transition rate, we simulate multiple systems with elastic properties parallel to that of real crystalline systems ($Y_0=20 \,  \text{eV Å}^{-2}$, $\kappa=1.2 \, \text{eV}$) at fixed length $L=20 a$ ($\approx 50\text{ Å}$ for graphene) and for a range of temperatures $k_BT/\kappa\approx 0.01-2$ (or $L/\ell_{th}\approx 0.8-5$) and aspect ratios $\alpha=2-9$ \footnote{Note that at fixed $L$, $\alpha$ is controlled by the clamped width $W_{clamp}$ which is equal to $W_0$ unless otherwise stated}. Once we ensure thermalization we can proceed to analyze the dynamics of the tagged node. We estimate the thermalized length of the system, $L_{th}$, as the length of the free membrane at a given temperature, projected onto the $z=0$ plane, shown in Fig.~\ref{sheet}(b). We then define the compressed length, $L_c$, as the projected length in the clamped configuration, illustrated in Fig.~\ref{sheet}(d).

The up-down transition rate is calculated by tabulating the average time spent in a tilted state, the dwell time $\tau_{dwell}$. Residence in the tilted state is determined conditionally with a threshold height $h_{th}=0.1\times L_0$: viz. $|h(t_n)|>h_{th}$ is assigned to the tilted state. The transition probability is then $\mathcal{R} \sim  1/\tau_{dwell}$.

A more elaborate method of {determining the dwell time}  is by computing the autocorrelation function of the time series post-thermalization. The normalized autocorrelation function, $\rho(\tau)\equiv C_t(\tau)/C_t(0)$, will decay exponentially with a time constant, $\tau_{ac}$. This time constant corresponds to the shortest time scale available {to} the system, which in this case is the time spent in {a given tilted state: $\tau_{dwell}\approx\tau_{ac}$}.

One can think of the dynamics of the {sheet} in the flat state, specifically the average height of the {tagged node, as a Brownian particle trapped in a harmonic well. In the pre-buckling regime the Langevin equation for the position $z(t)$ of a particle of mass $m$ is}
    \begin{equation}
        \ddot{z}(t) = -\frac{\gamma}{m}\dot{z}(t)-\omega_0^2 z(t) + \frac{1}{m}\xi(t)
        \label{h_langevin}
    \end{equation}
where $\omega_0=k/m$ and $\xi(t)$ is Gaussian noise with
    \begin{equation}
        \langle\xi(t)\rangle=0,\qquad\langle\xi(t)\xi(t')\rangle=2m\beta k_BT\delta(t-t').
        \label{noise_moments}
    \end{equation}
{Fourier transforming ($\dv[n]{}{t}z(t)\rightarrow (-i\omega)^nz(\omega)$) gives}
    \begin{equation}
        z(\omega) = \frac{1}{m}\frac{\xi(\omega)}{\omega_0^2-\omega^2+i\frac{\gamma}{m}\omega}.
        \label{h_omega_sol}
    \end{equation}

\begin{figure}[t]
        \centering 
        \def\svgwidth{\columnwidth}
        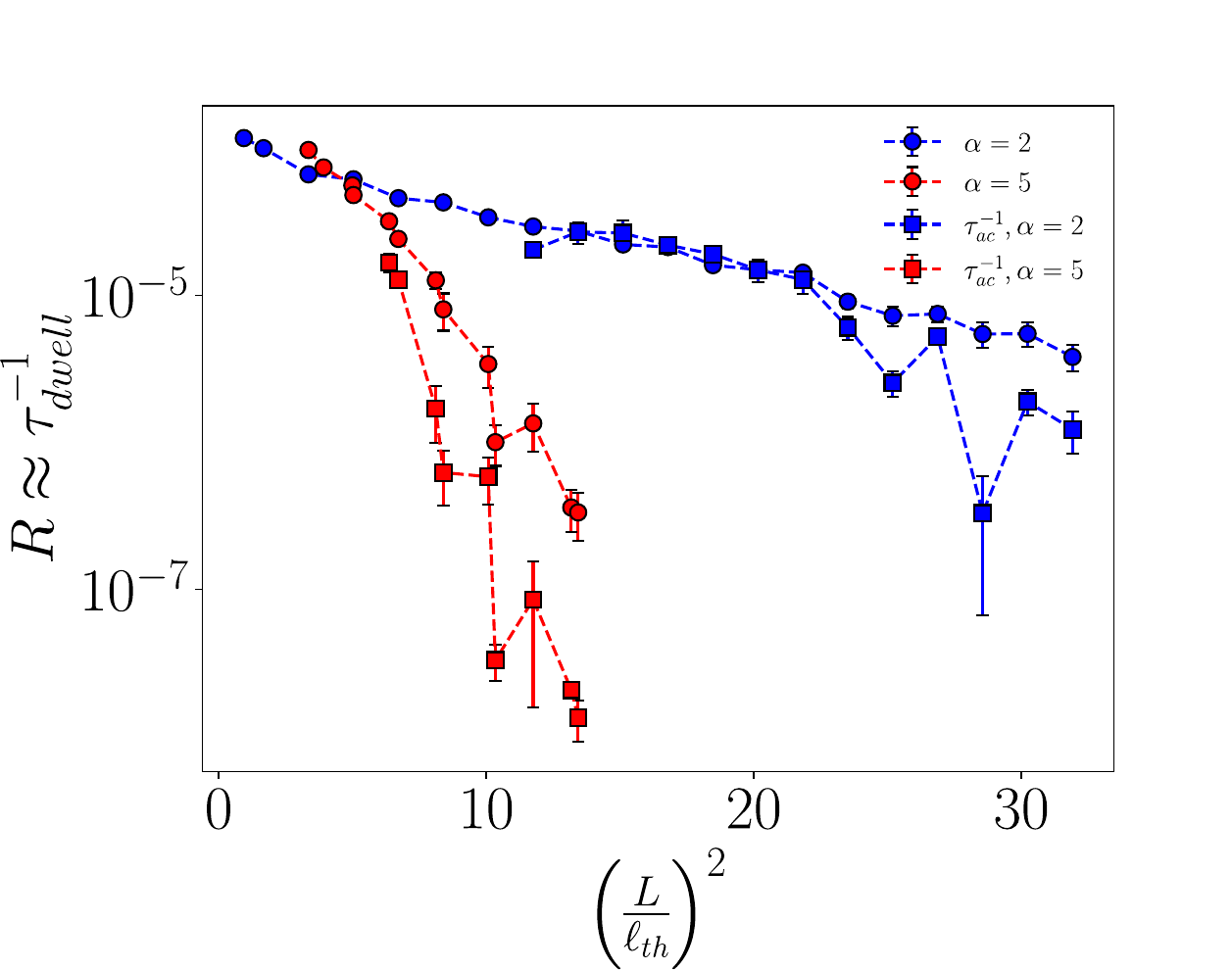
        \caption{Transition rates for $\alpha=2,5$ as a function of temperature as measured by $\frac{L}{\ell_{th}}$. The points are computed from both dwell time averaging over multiple runs  (circles) and from extracting the time constant $\tau_{ac}$ from a fit of the autocorrelation function, $\rho(\tau)$, to Eqs.(\ref{rho_tau},\ref{rho_scaling}) for the regime with observable transitions and the fully tilted, low transition-rate regime, respectively (squares). }
        \label{corrcomp}
\end{figure}

We can now compute the correlation function via {the} inverse Fourier transform of the squared average in frequency space,
    \begin{align}
        C_t(t') = \langle z(t)z(t')\rangle &= \int_{-\infty}^{\infty}\frac{d\omega}{2\pi}\,\langle
        |z(\omega)|^2\rangle e^{-i\omega t} \\
        &= \frac{\gamma k_BT}{\pi m^2}\int_{-\infty}^{\infty}d\omega\,\frac{e^{-i\omega t}}{(\omega^2-\omega_0^2)^2+\frac{\gamma}{m}\omega^2}.
        \label{C_t_int}
    \end{align}
{The integral in Eq.~(\ref{C_t_int}) via complex methods with a semicircular contour}
    \begin{equation}
        C_t(\tau) = \frac{k_BT}{m\omega_0^2}e^{-\frac{\gamma \tau}{2m}}\left[\cos\omega_1 \tau+\frac{\gamma}{2m\omega_1}\sin\omega_1 \tau\right].
        \label{C_t_sol}
    \end{equation}
The normalized autocorrelation function is then
    \begin{equation}
        \rho(\tau) = \frac{C_t(\tau)}{C_t(0)} = e^{-\frac{\gamma \tau}{2m}}\left[\cos\omega_1 \tau+\frac{\gamma}{2m\omega_1}\sin\omega_1 \tau\right].
        \label{rho_tau}
    \end{equation}
For sufficiently long times ($\tau\gg t_{dwell}$) and for systems with low tilt transitions the autocorrelation will decay as \cite{amit2005field} 
    \begin{equation}
        \rho(\tau) \sim \exp(-\tau/\tau_{ac}).
        \label{rho_scaling}
    \end{equation}
{Comparing Eqs.~(\ref{rho_tau}) and (\ref{rho_scaling}) shows that}
    \begin{equation}
        \tau_{ac} \approx \frac{2m}{\gamma}
        \label{tau_ac}
    \end{equation}

We now provide a more detailed comparison between the simple average dwell time method and the autocorrelation method.

We compute the autocorrelation function of the height time series within a state and fit the curve to a function of the form of Eq.~(\ref{rho_tau}), extracting the time constant, $\tau$. We can then compare to our previous results. Fig.~\ref{corrcomp} shows a semilog plot of $\tau_{\rm dwell}^{-1}$ as a function of $\left(L/\ell_{th}\right)^2$, which is proportional to $k_B T$. The transition rate data $\tau_{\rm dwell}^{-1}$ are approximated using two methods: (i) height filtering and (ii) fitting to the autocorrelation function.  Note that in the tilted regime, using autocorrelation to extract the time constant may sample smaller timescales than the one of interest, namely the transition time, $\tau_{trans}<\tau_{dwell}$, which is the time it takes to jump from one state to the other. Fig. ~\ref{corrcomp} shows autocorrelation estimates and we see that both methods provide roughly the same probability values and trends. {There is, however, high variance} in the autocorrelation estimate which can be attributed to uncertainty in the fit. The two methods agree qualitatively and for the purposes of studying the trends in the transition rate, {we chose the former to save computational time.}

\begin{figure}[ht]
        \centering 
        \def\svgwidth{\columnwidth}
        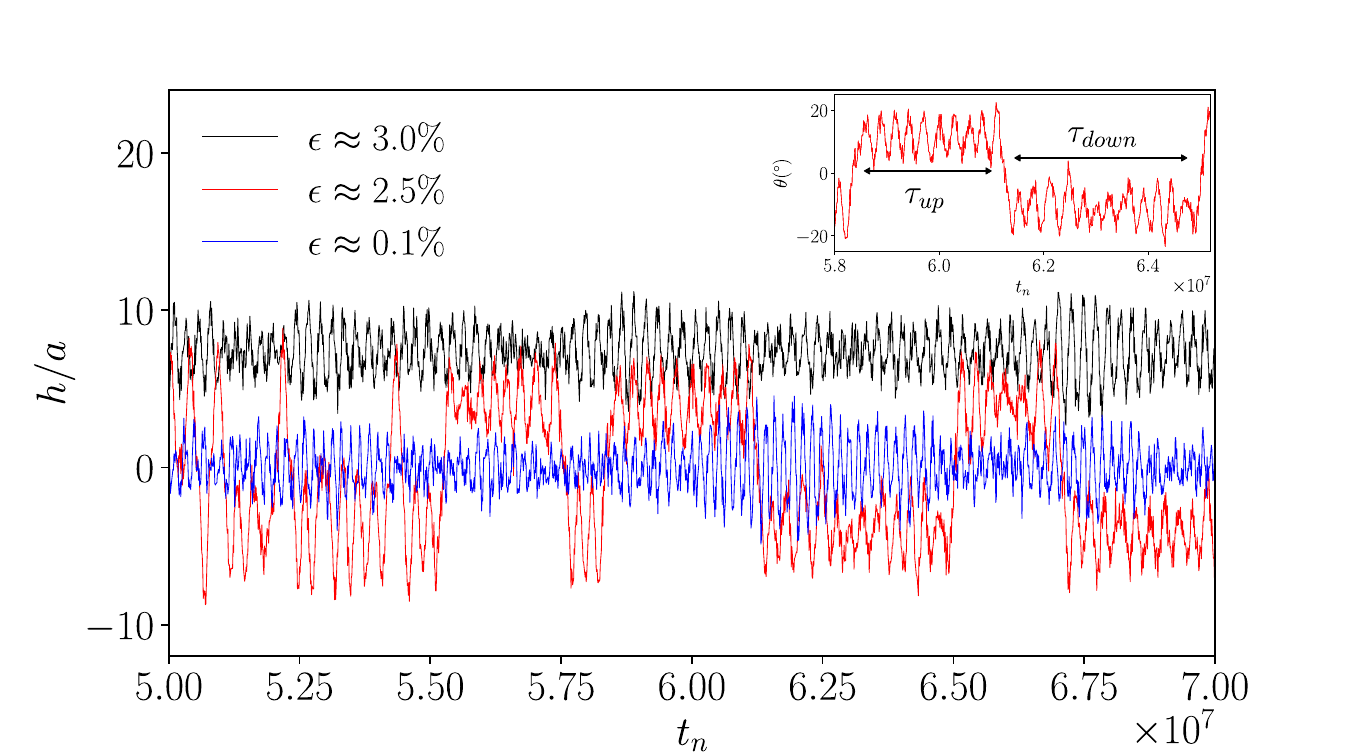
        \qquad
        \def\svgwidth{\columnwidth}
        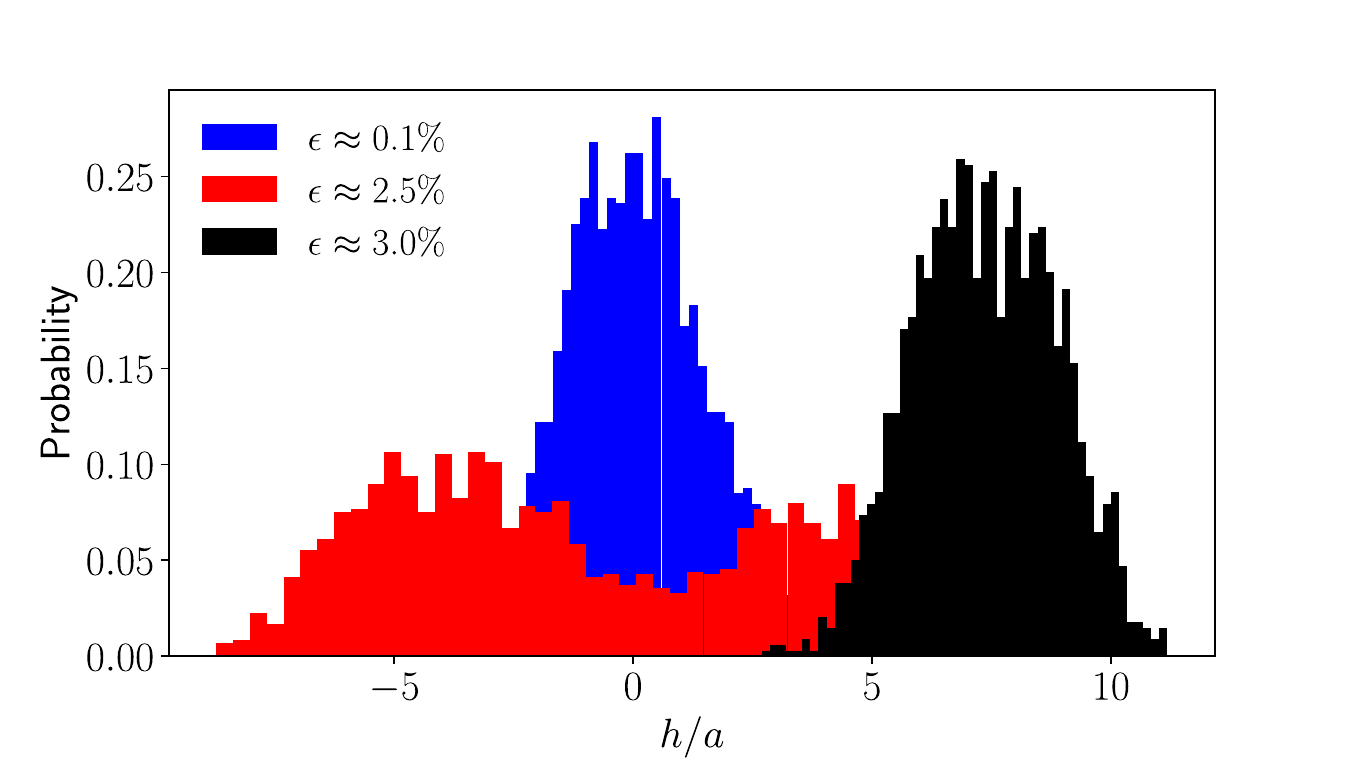
        \caption{(top) Simulated time series of the height, $h(t_n)$, at timestep $t_n$  of the middle slice at the free end of a sheet for $\alpha=3$ and $\kappa/k_BT = 4$ ($L/\ell_{th}
        \approx3$). Each curve represents a different value of the clamp strain, $\epsilon = (W_{clamp}-W_{th})/W_{th}$ (in \%). Note that the height can be easily translated into a tilt angle: see the red curve which oscillates between $\theta=\pm 20^\circ$, as measured from the $xy$ plane (see inset). Inset shows examples of the definition of the up and down dwell times, $\tau_{up/down}$. (Bottom) Histogram of the probability distribution of heights for the time window of the time series above. Each distribution corresponds to a different clamping strain and shows the three possible cases: unimodal with zero mean (blue), bimodal (red), and unimodal with nonzero mean. These correspond to the states where the sheet is flat, tilted with frequent transitions, and tilted with infrequent transitions, respectively. }
        \label{timeser}
\end{figure}

\subsubsection{Clamping} 

 To investigate the role of clamping we simulated several systems clamped at a range of strains close to  $W_{clamp}= W_{th}$. Fig. \ref{timeser} shows the time series of the height field $h=h(x=L,y=0)$, for several clamping strains, $\epsilon\equiv(W_{clamp}-W_{th})/W_{th}$. Clamping sufficiently close to $W_{th}$ does not induce tilt -- the sheet fluctuates  about a mean horizontal state (blue curve).  Above a positive parameter-dependent threshold for $\epsilon$ we see the onset of tilt and the accompanying up-down inversions (red curve). The dwell time increases with $\epsilon$ (black curve). 

{It is instructive to measure the effective plane stress throughout the clamped sheet, as determined by displacements with respect to a fixed average thermalized free state.} We take the particle positions of a sheet configuration at a fixed timestep and compute the displacements relative to the average thermalized, free reference state. Treating our lattice as a triangulated mesh and embedding the displacement field to the vertices of this mesh, we compute the linear 2D plane stress via finite element analysis. Appendix A describes a related method that transforms the stresses back into the nodal basis.

\begin{figure}[ht]
        \centering 
        \def\svgwidth{.8\columnwidth}
        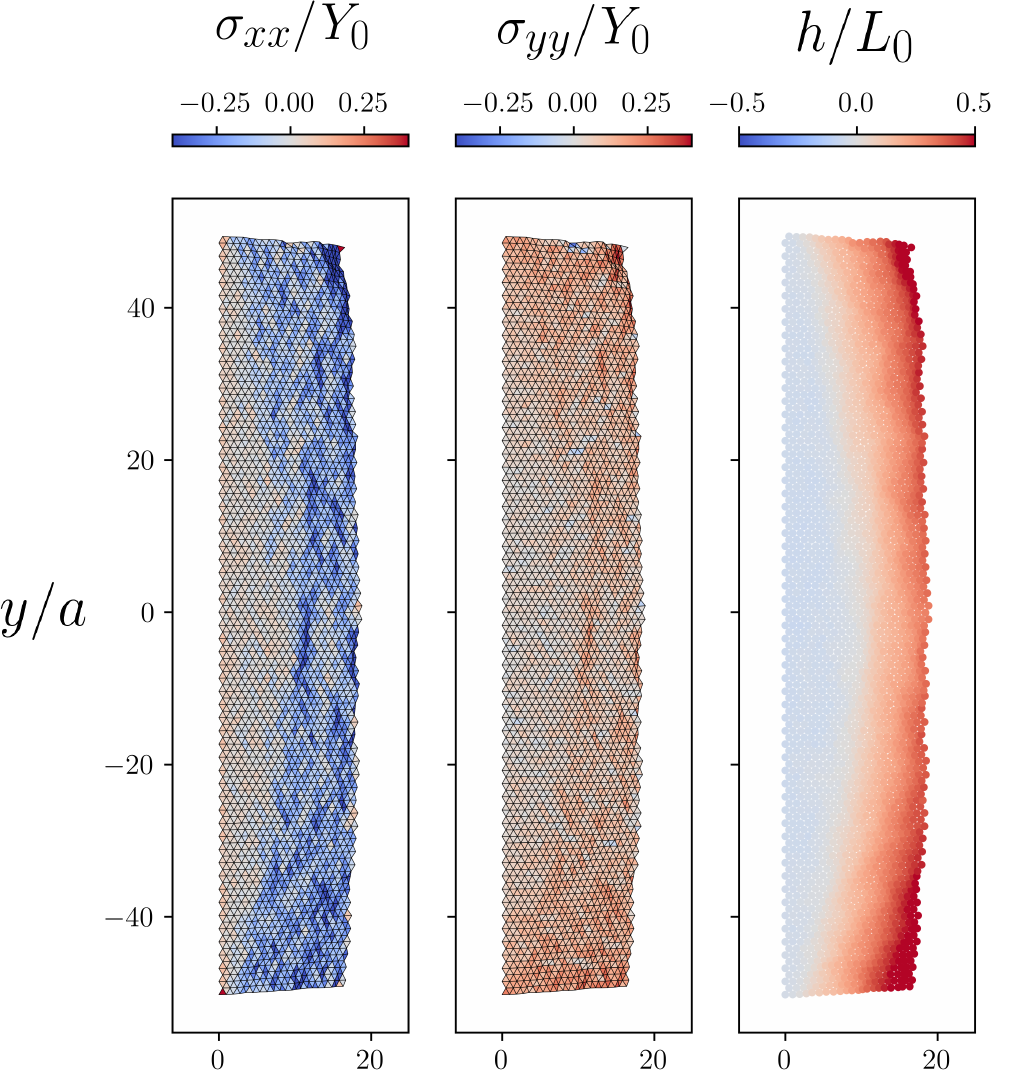
        \qquad
        \def\svgwidth{.8\columnwidth}
        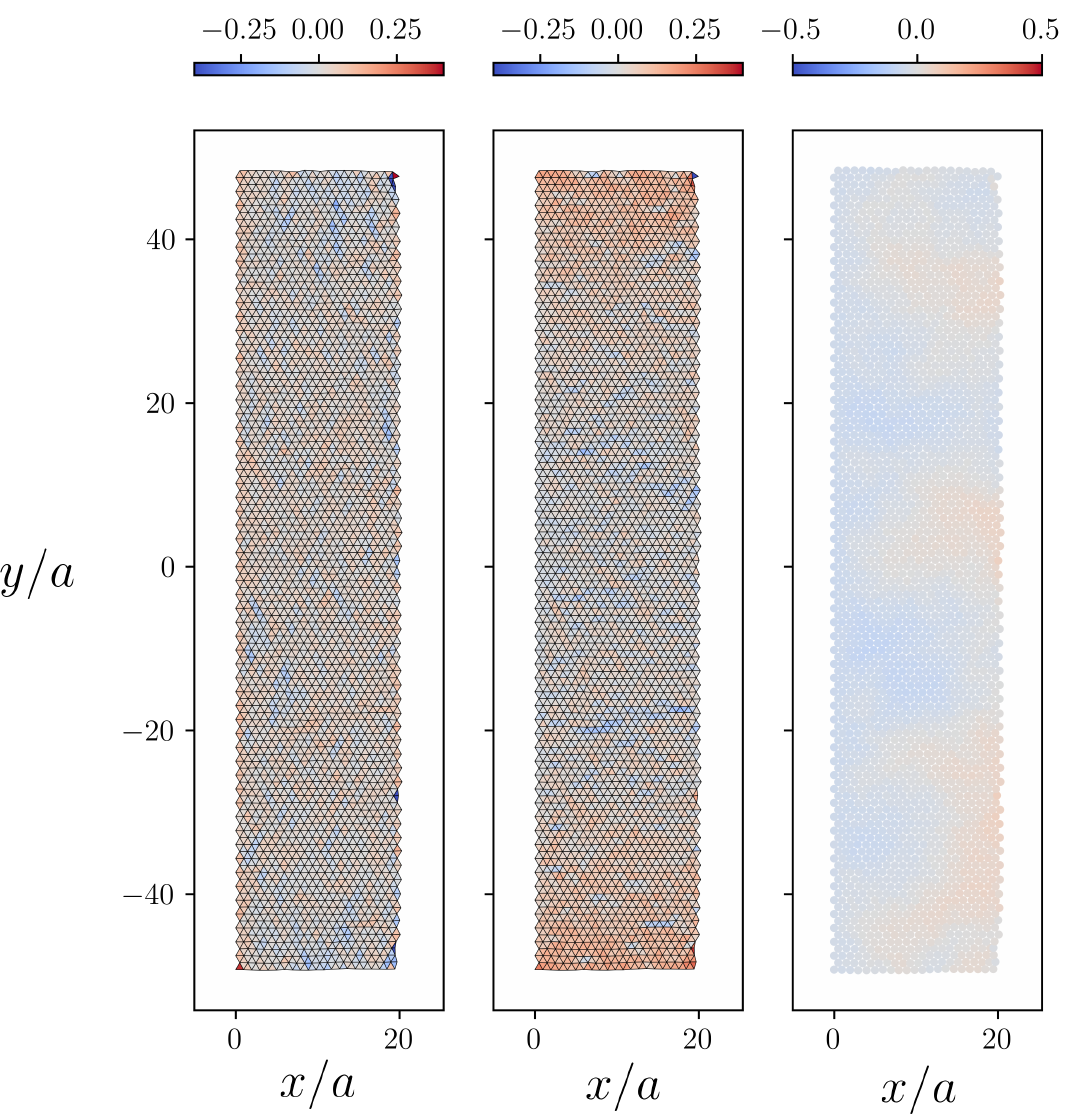
        \caption{Diagonal components of the effective plane stress on the simulated sheet with $\alpha=5$ and $L/\ell_{th}\approx3$ along with a map of the height field for (top) $W_{clamp}=W_0>W_{th}$ and (bottom) $W_{clamp}\approx W_{th}$. Sheet is shown as the projection of the deformed state onto the plane. The strain tensor is computed directly by differentiating the displacements from the free thermalized state.}
        \label{FEMstress}
\end{figure}

For an effective extensive strain concentrated at the clamp ($W_{clamp}>W_{th}$), there are two competing effects: (1) the zero-temperature elastic response associated with a standard positive Poisson ratio, leading to compression along $x$ and (2) the response associated with a negative Poisson ratio thermalized sheet associated with the known behavior at the thermal Foppl-von Karman fixed point, which creates an extensive response along $x$. The first effect should dominate in a zone of influence near the clamp as stretching suppresses thermal fluctuations. The second effect should dominate sufficiently far from the clamp where the sheet closely resembles a free fluctuating membrane. Fig. \ref{FEMstress} shows a simulated map of both diagonal elements of the stress tensor at fixed time steps obtained using the finite element method described above. We see that the $\sigma_{xx}$ component confirms compressive stress for the tilted state ($W_{clamp}=W_0>W_{th}$) and very little stress in the flat state ($W_{clamp}\approx W_{th}$), as expected. The $\sigma_{yy}$ component shows the expected extension in the tilted phase. We note that the flat phase also exhibits some extension, which we attribute to the fact that we have a nearly but not quite zero strain at the clamp. We can also compare these stress maps to results found in previous work on tilted flaps \cite{chen_spontaneous_2022}. A detailed comparison can be found in Appendix B.

\begin{figure}[ht]
        \centering 
        \def\svgwidth{\columnwidth}
        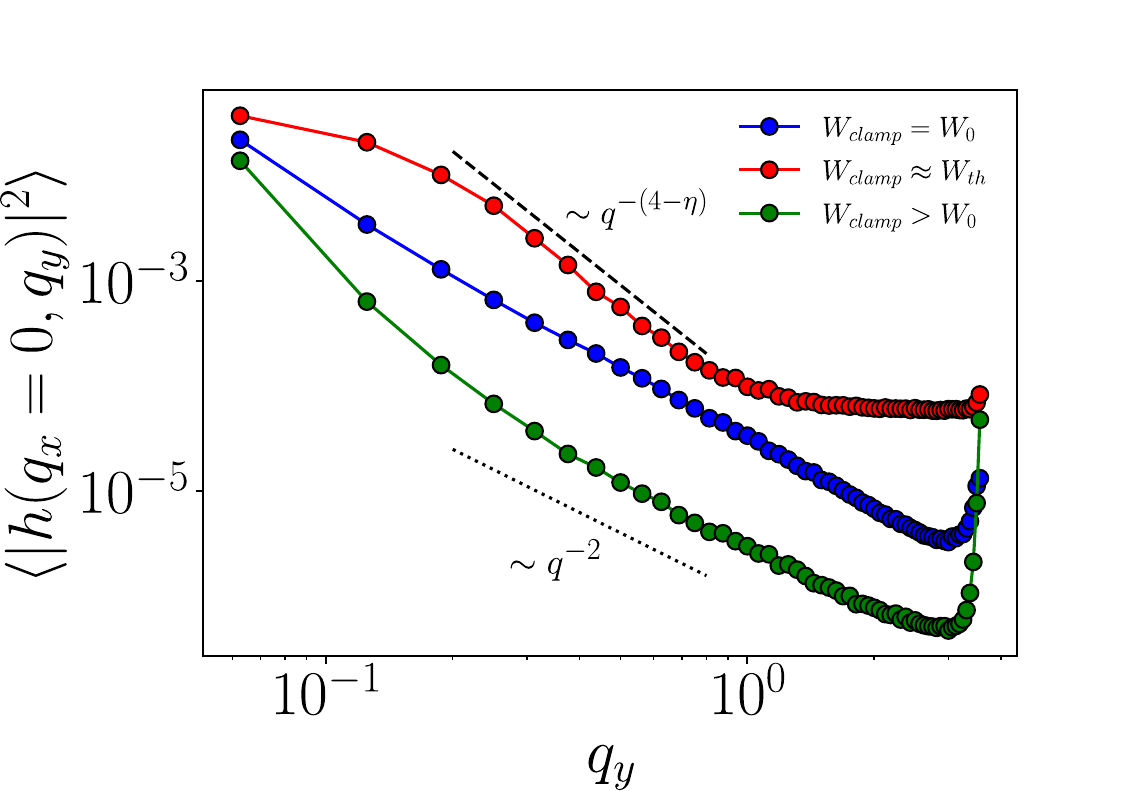
        \caption{Height-height spatial correlation function, $\langle|h(q_x=0,q_y)|^2\rangle$, as a function of wavevector in the $y$-direction at fixed temperature $L/\ell_{th}\approx3$. Clamping near the thermalized width, $W_{th}$ (red, $\epsilon\approx0.007$), leads to a scaling with power law $q^{-(4-\eta)}$ indicating a negligible stress term in Eq.~\ref{two_point_corr}. Note that $W_{th}<W_0$. For non-zero strain, we have $W_{clamp}=W_{0}, \epsilon\approx0.013$ (blue) and $W_{clamp}>W_0, \epsilon\approx 0.024$ (green). The correlations have quadratic scaling, indicating there is significant stress in the $y$ direction.}
        \label{corr}
\end{figure} 

One can confirm that the flat state reflects a {zero-stress configuration from by the height-height correlation function, which is expected to scale as}
    \begin{equation}
        \langle |h(q)|^2\rangle \approx \frac{k_BT}{A(\kappa_R(q) q^{4}+\sigma_{ij} q_iq_j)}.
        \label{two_point_corr}
    \end{equation}
where $\sigma$ is the stress due to the clamp, $A$ is the projected area of the sheet and $\kappa_R$ scales as in Eq.~\ref{kappa_R}. In the absence of stress, the bending term dominates and the correlation function will scale as $q^{-(4-\eta)}$. On the other hand, if there is a significant source of stress, we expect the quadratic term to dominate. Fig.~\ref{corr} shows the mean-squared height fluctuations in momentum space for three classes of clamp width. For clamping near the thermalized width ($\varepsilon\approx0$) the slope is approximately $ -(4-\eta)$, indicative of bending dominance.  The other two classes exhibit  a quadratic fall-off, indicating stress dominance at low wavevector (see Eq.~\ref{two_point_corr}).



\begin{figure}[ht]
        \centering 
        \def\svgwidth{\columnwidth}
        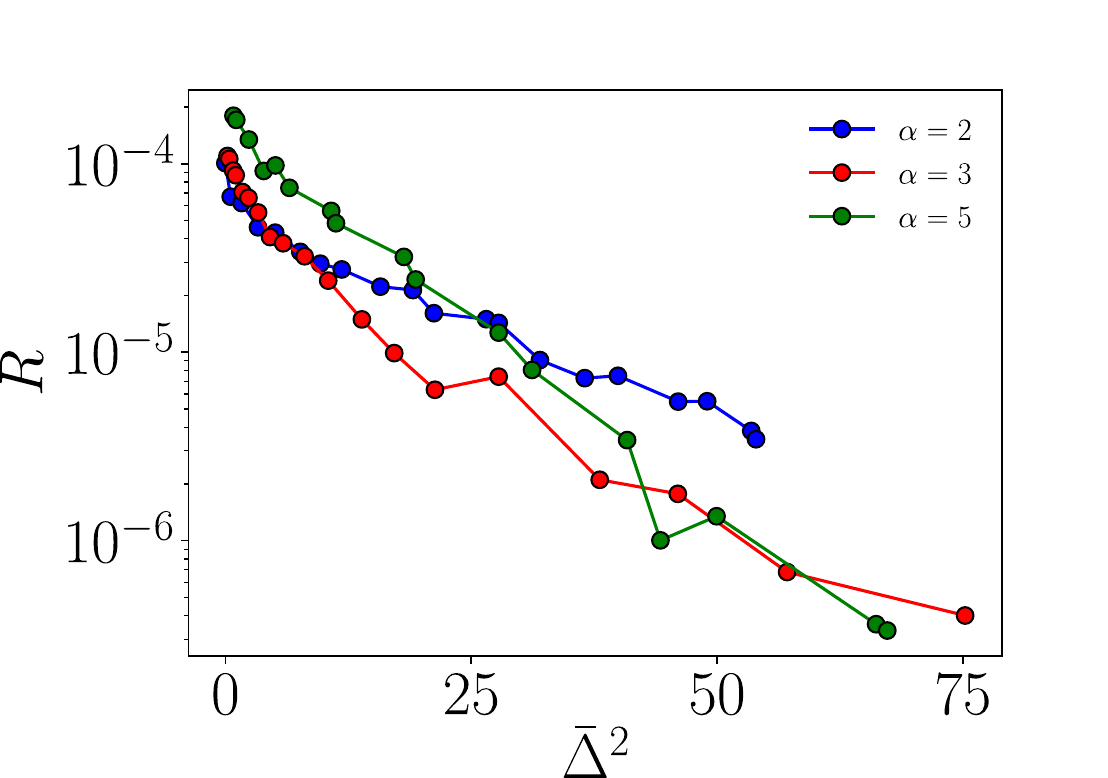
        \qquad
        \def\svgwidth{\columnwidth}
        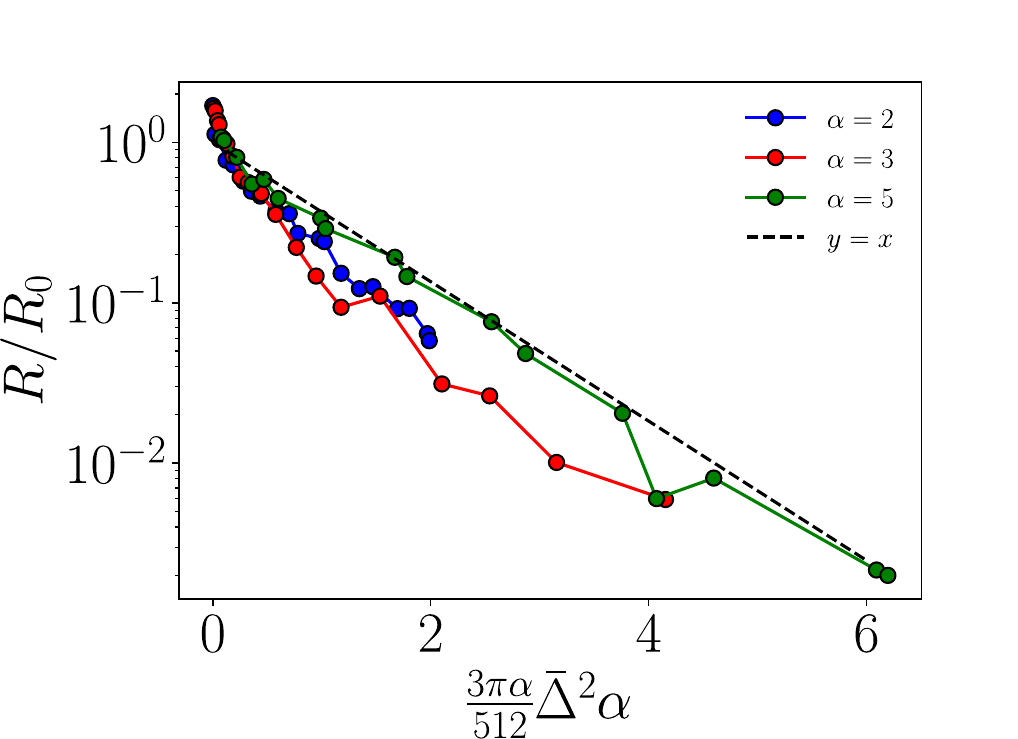
        \caption{(Top) Transition rate as a function of the squared relative compression, $\bar\Delta$. Each curve represents a different value of the aspect ratio. (Bottom) Same data as the plot above but scaled by the $R_0$ obtained from fitting the top curves to a line. The $x$ axis is scaled by the constants predicted in Eq.~$\ref{R_therm}$. We see a near collapse onto the line $y=x$, as predicted for a temperature independent Arrhenius factor.}
        \label{rvsarg}
\end{figure}

\subsubsection{Comparison to Kramers' theory}

We proceed to compare our molecular dynamics simulations to the predictions made by the underdamped Kramers' theory along with the elastic mean field theory described in Sec. I. We first estimate $\bar\Delta=(\Delta-\Delta_c)/\Delta_c$ using a time average of the in-plane displacement at a variety of temperatures. The critical compression is obtained by computing the height susceptibility in analogy to the classical Ising model~\cite{hanakata_thermal_2021}.

Eq.~\ref{R_therm} predicts that the log of the transition rate falls linearly with slope $-3\pi \alpha \bar\Delta^2/512$. Fig.~\ref{rvsarg} shows a semi-log plot of the transition rate as a function of $\bar\Delta$: the curves are indeed linear at large $\bar\Delta$. Further confirmation of ~\ref{R_therm} is found by normalizing $R$ by the best-fit $y$-intercept $R_0$. The bottom plot of Fig. ~\ref{rvsarg} shows the normalized transition rate, $R/R_0$, as a function of the full argument $3\pi \alpha \bar\Delta^2/512$. We see a near linear collapse.

Thermal fluctuations usually promote transitions between distinct energy minima. We find instead that they suppress transitions in the $L\gg \ell_{th}$ regime, locking the membrane in one of the two tilted states. Recall that displacements are being measured with respect to a free-standing configuration where thermal fluctations shrink the overall area: $W(T_1)<W(T_2)$ for $T_1>T_2$. As temperature is increased the strain induced at the clamp grows, driving the system deeper into the tilted phase.

\subsubsection{The role of geometry}

\begin{figure}[t]
        \centering 
        \def\svgwidth{\columnwidth}
        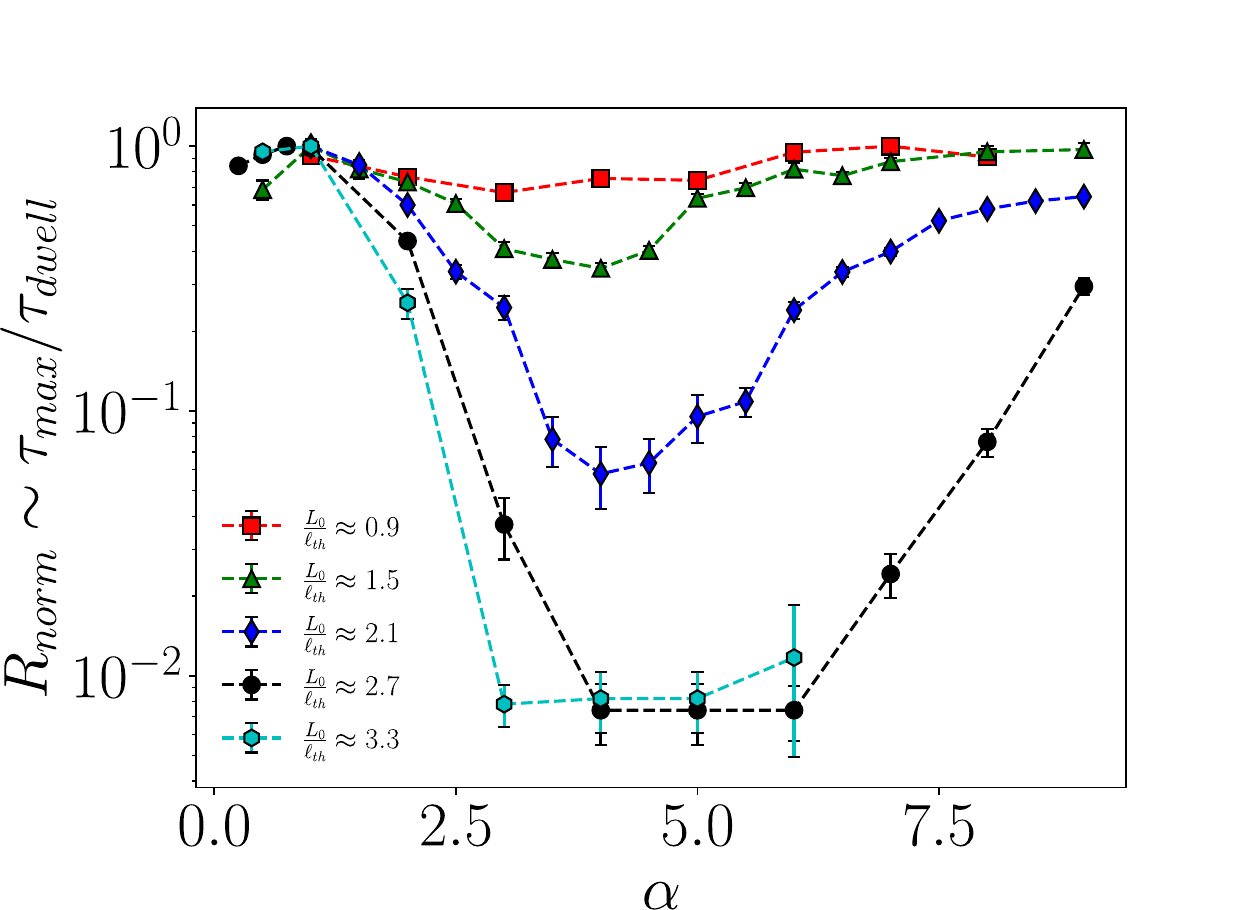
        \caption{Inverse average dwell time as a function of the aspect ratio, $\alpha$, for various temperatures as shown in the legend. We see a clear dependence on $\alpha$, with a peak reaching nearly the entire simulation time around $\alpha=4-5$, indicating the sheet is tilted for the entire simulation run. Note that for $L<\ell_{th}$ (red points), while the dwell time is non-zero, it is significantly lower than the thermalized curves.}
        \label{rvsalpha}
\end{figure}

We now turn to the role of geometry as controlled by the aspect ratio. Geometrical tuning offers a very different addition to the experimental toolkit which may well be more feasible and reliable~\cite{zande2010large, masih2016controlled} than precise tuning of temperature and does not require any new materials or external fields. To explore this dependence we fix a temperature in the tilted phase and simulate a set of distinct aspect ratios in the range $0.5<\alpha<9$ and extract the simulated dwell time, $\tau_{dwell}$. Fig.~\ref{rvsalpha} shows the inverse dwell time normalized by the maximum value for a set of  temperatures corresponding to $L/\ell_{th}\approx 0.9,2.1,2.7,3.3$. For $L>\ell_{th}$, there is a clear $\alpha$-dependence with a minimum for  $\alpha\approx4-5$. We can compare this to Fig.~\ref{phaseplot}(b) where Kramers' theory also predicts this low transition-rate region. This $\alpha$-window corresponds to confinement in one of the tilted wells with rare transitions.

\section{Conclusion}
Combining a one-dimensional mean-field model of a thermalized thin elastic sheet with cantilever boundary conditions and Kramers' transition state theory, we have analyzed the transition dynamics of the tilted state in the regime where the width exceeds the length ($\alpha>1$). Renormalization of the elastic constants due to thermal fluctuations beyond the thermal lengthscale leads to a cancellation of temperature in the Boltzmann factor of the transition probability, leaving a dominant dependence on the aspect ratio. Implicit temperature dependence enters via the relative compression $\bar{\Delta}$, slowing the dynamics and suppressing transitions between the two degenerate tilted states. Below the critical crumpling transition, the transition rate is low, locking the system in one of the two tilted phases. A key role is played by the effective stress at the clamp with respect to a free thermalized sheet.

The predictions of Kramers theory are verified by analyzing the variation of the transition rate with the compression $\bar\Delta$. The transition rate dependence in the transition rate exhibits the expected Arrhenius behavior $\sim \exp(-C\bar\Delta^2)$ with $C=3\pi\alpha/512$.

Clamped thermalized sheets possess a rich dependence on the purely geometrical aspect ratio with the transition rate reaching a minimum for $\alpha_{min}\approx 4$.

\begin{figure}[t]
        \centering 
        \def\svgwidth{\columnwidth}
        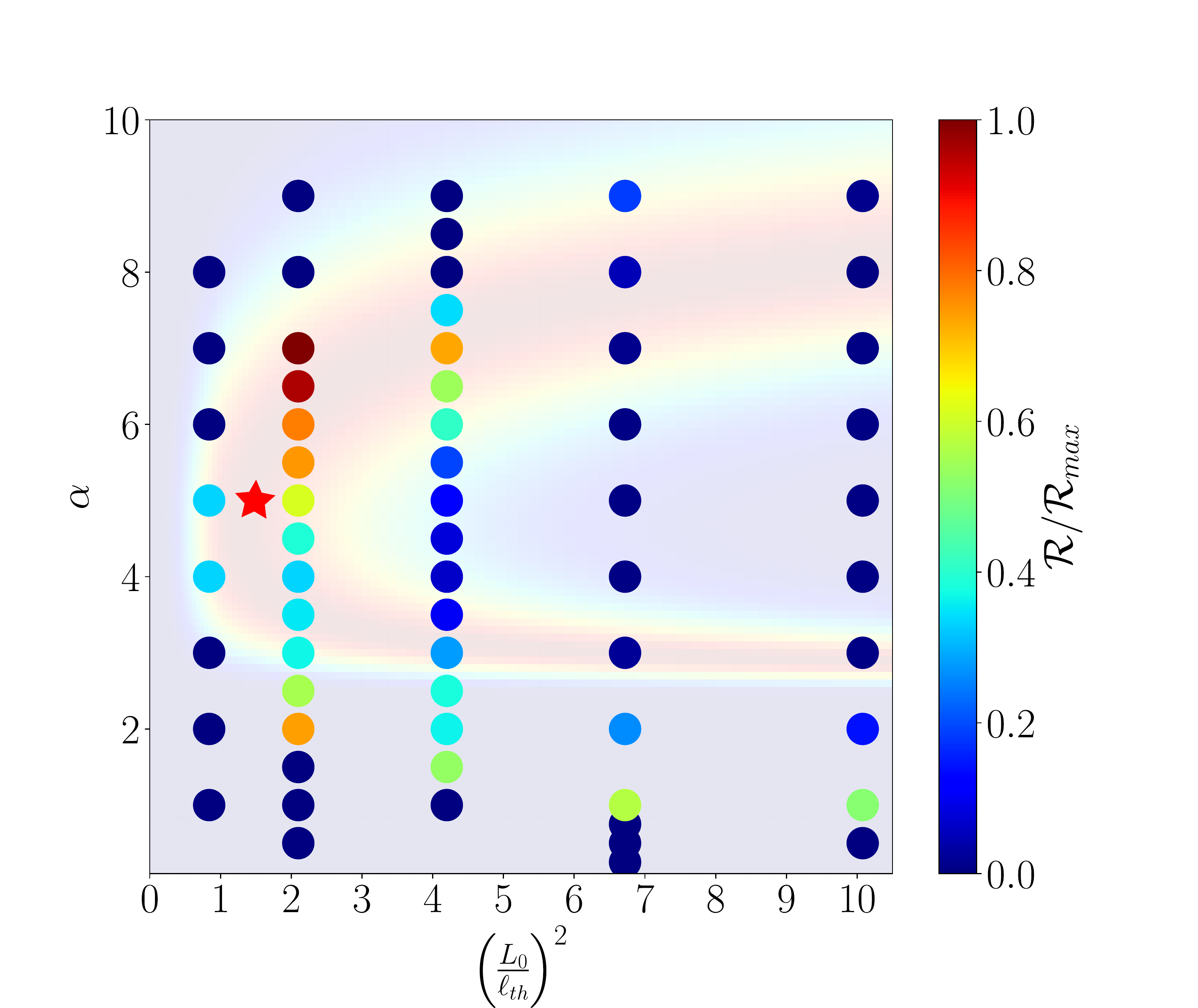
        \caption{Density plot of the transition rate as a function temperature and aspect ratio as obtained from molecular dynamics data. Comparing this plot to Fig.~\ref{phaseplot} (shown as the background colors) we  see general agreement with the high rates (red) localized in two regions of $\alpha$, separated by a dynamically stable region (blue). For reference, the red star in the plot denotes a graphene sample of the same size ($L\approx50\text{ Å}$) at room temperature.}
        \label{densdata}
\end{figure}
The temperature ranges where we observe the behavior studied here are currently beyond standard 2D-metamaterials such as micron-scale room temperature graphene. Perforated sheets and other kirigami-like structures \cite{yllanes_thermal_2017} which lower the bare bending rigidity and enhance bending fluctuations, as well as permitting new bending configurations, may allow the observation of tilt and its dynamics in experimentally realizable systems. 
In particular geometric control of the dynamical switching exhibited by the elastic sheets studied here should have rich applications in micro- and nanoelectromechanical systems (MEMS/NEMS) \cite{xu_nanomechanical_2022, lifshitz_nonlinear_2008}.

\section{Acknowledgements}
This research was supported in part by the National Science Foundation under Grant No. NSF PHY-1748958. This material is based upon work supported by the National Science Foundation California LSAMP Bridge to the Doctorate Fellowship under Grant No. HRD-1701365. P.Z.H acknowledges support through NSF Grant No. DMR-1608501 and via the Harvard Materials Science Research and Engineering Center, through NSF Grant No. DMR-2011754. We also thank the KITP program, ``The Physics of Elastic Films: From Biological Membranes to Extreme Mechanics,'' supported in part by the National Science Foundation under Grant No. NSF PHY-1748958.

\appendix    

\begin{figure}[ht]
        \centering 
        \def\svgwidth{.9\columnwidth}
        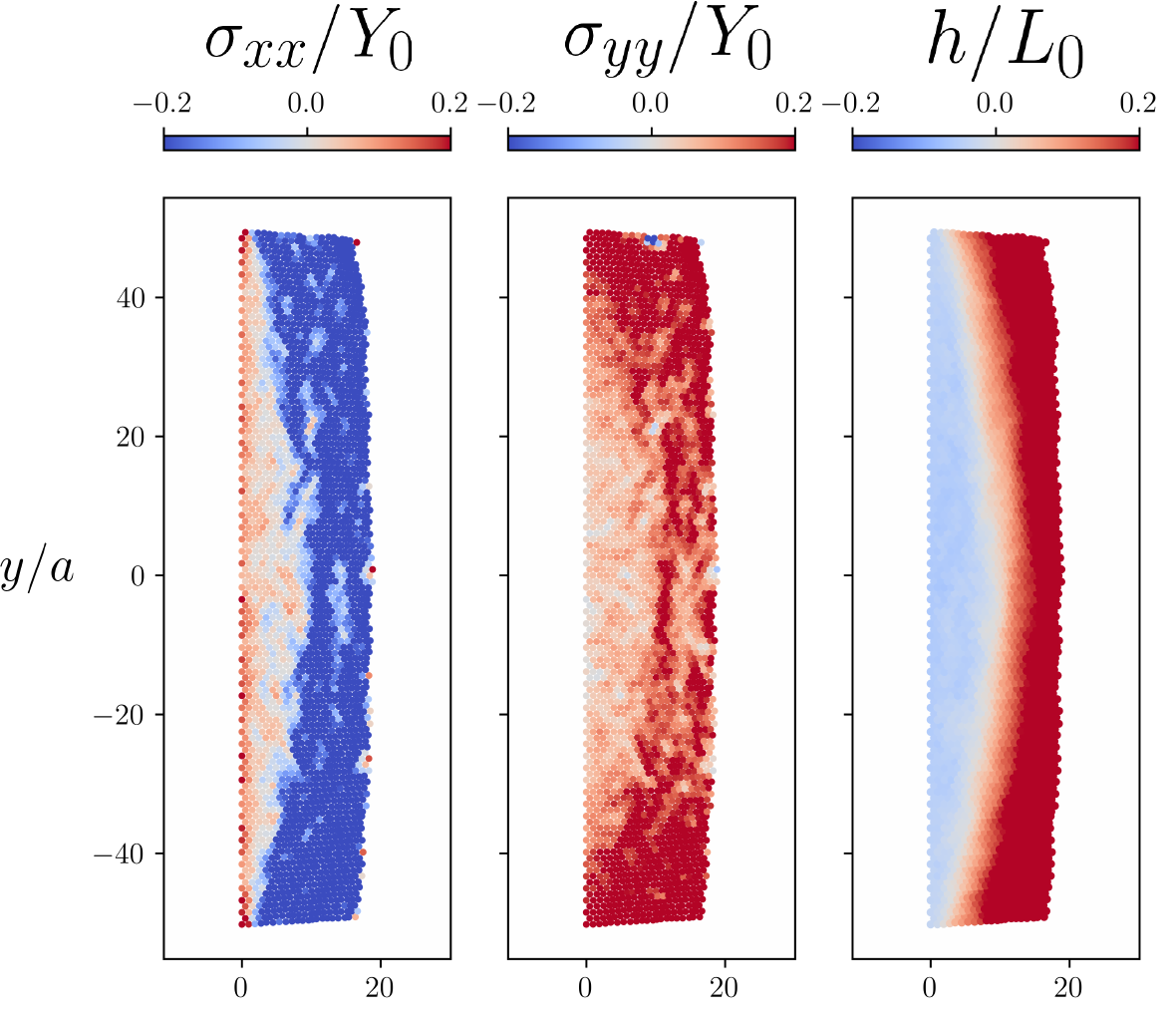
        \qquad
        \def\svgwidth{.9\columnwidth}
        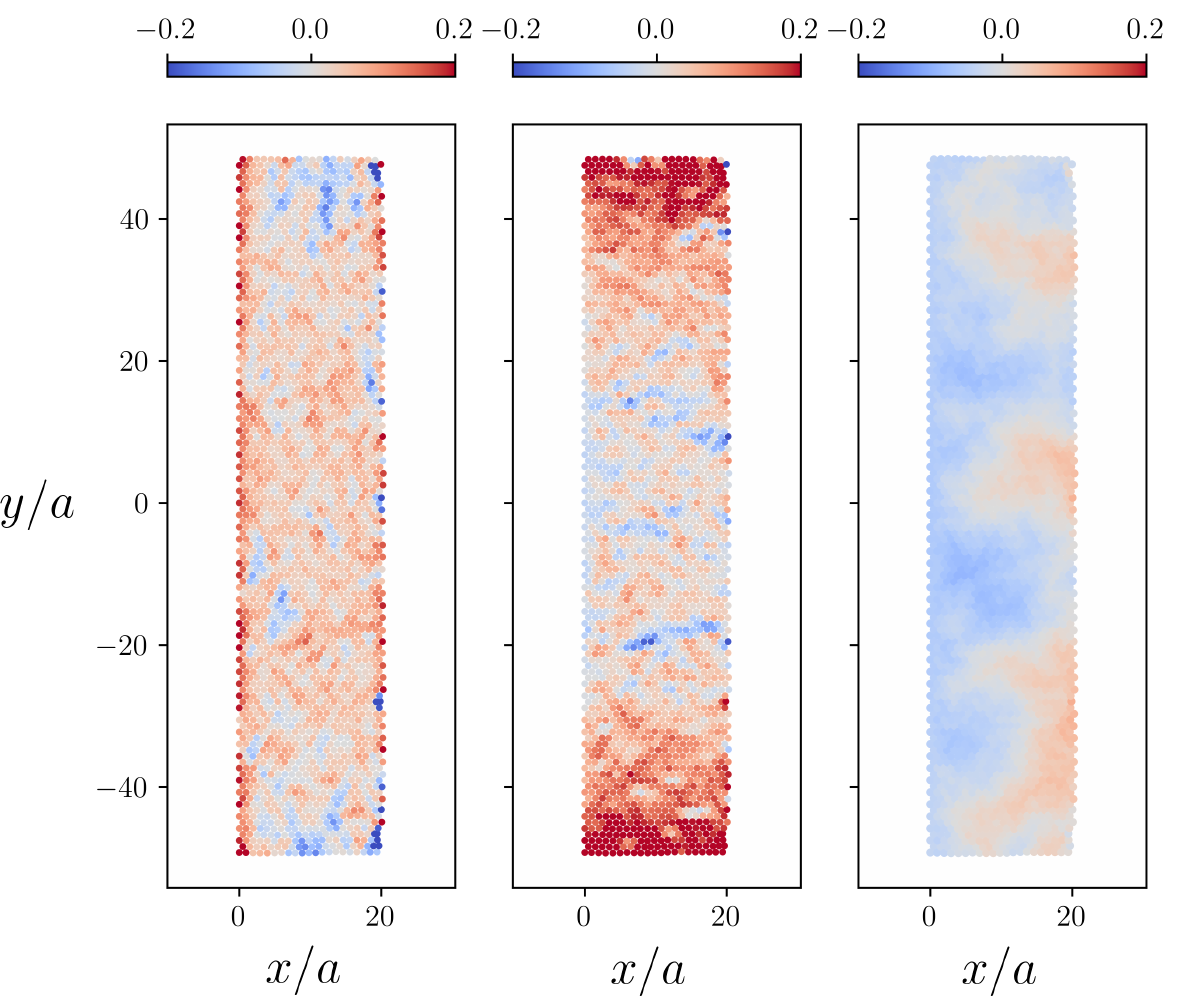
        \caption{Diagonal components of the effective plane stress on the simulated sheet along with a map of the height field for (top) $W_{clamp}=W_0>W_{th}$ and (bottom) $W_{clamp}\approx W_{th}$. Strain tensor is computed directly by differentiating the displacements from the free thermalized state. }
        \label{gradavg}
\end{figure}

\begin{figure}[ht]
        \centering 
        \def\svgwidth{.7\columnwidth}
        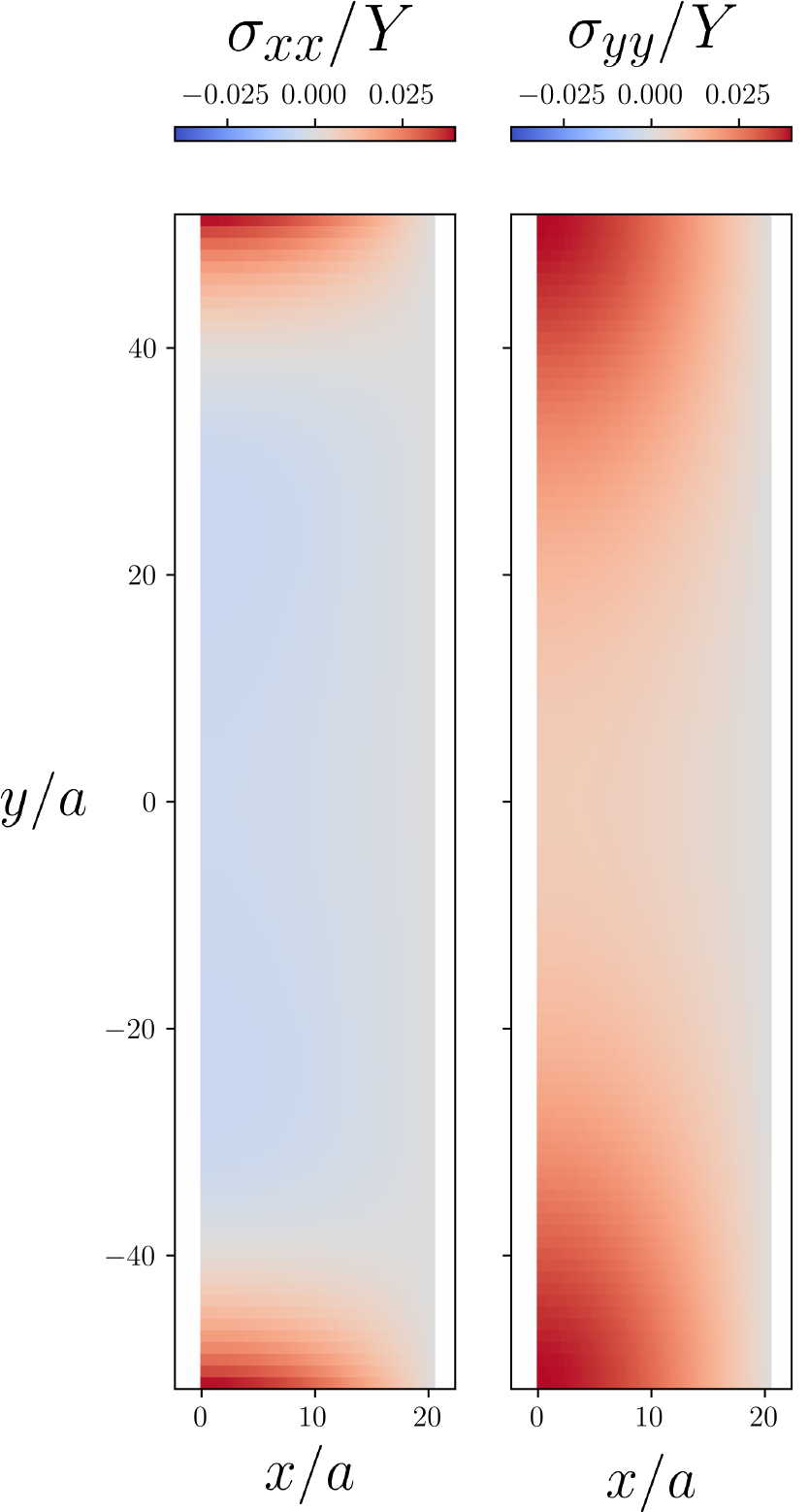
        \caption{Theoretical approximation of the plane stress for an $\alpha=5$ sheet with extension $\epsilon=0.02$ at $T=0$.}
        \label{sigma_theory}
\end{figure}

\section{Gradient average estimate of stress tensor}

An alternative method to obtain a version of Fig.~\ref{FEMstress} is via discrete gradient averaging of the gradient operator on the lattice. This is a method that is commonly implemented on various differential operators defined on meshes \cite{MANCINELLI201937, mesh}. 

Consider a graph $G=\{V,E\}$ with vertices $V$ coinciding with the vertices $V$ and edges $E$ of our simulated lattice and we define some vector field $\vec{f}_i: V\rightarrow \mathbb{R}^2$, for $i\in V$. We start by focusing on a single vertex $v_i\in V$ and its neighborhood (or star) $\mathcal{N}(v_i) = \{T_I\}_{I=1}^6$ where $T_I$ is 1 of the 6 triangles making up the neighborhood of $v_i$. We then compute the gradient defined in each triangle $T_I$. This is done via barycentric interpolation of the three values of $f_i$ on the vertices $T_I$. The gradient at triangle $T_I$ is given by 
    \begin{equation*}
        (\nabla f_{T_I})_{ij} = (f_i-f_k)\frac{(v_k-v_j)^\perp}{2 A_{T_I}} + (f_j-f_k)\frac{(v_i-v_k)^\perp}{2A_{T_I}}
    \end{equation*}
where $v^\perp$ is the $90^\circ$ rotation of vector $v$ and $A_{T_I}$ is the area of triangle $T_I$. This gives a gradient tensor in the basis of the faces of $G$.

In order to move back to the basis of vertices we compute a weighted average over the neighborhood of vertex $v$ and define the gradient at $v$ as 
    \begin{equation*}
        (\nabla f_{v})_{ij} = \frac{1}{\sum_{T_I\in \mathcal{N}(v)} A_{T_I}}\sum_{T_I\in \mathcal{N}(v)}A_{T_I}(\nabla f_{T_I})_{ij}
    \end{equation*}

We can now compute the mesh gradient of the in-plane displacement field, $u_i$, as defined on the vertices of the deformed lattice. The in-plane gradient is then obtained by symmetrizing the gradient as computed above, that is, $U_{ij} = (\partial_iu_j +\partial_j u_i)/2$. Fig.~\ref{gradavg} below shows the strain map obtained using this method. Note that this gives very similar results to the method shown in the main text.

\section{Comparison of Clamping stress to theory}

Previous work \cite{chen_spontaneous_2022} on this system estimated the in plane stress using a doubling method that converts the clamped boundary condition into an internal condition. This results in the stress components shown in Fig.~\ref{sigma_theory} for a sheet of $\alpha = 5$ and $\epsilon = 0.02$. Comparing to the simulated stresses (Figs.~\ref{FEMstress},\ref{gradavg}) we see that we have a similar accumulation of high extensive stress in the $yy$ component for both theory and simulation, indicating the extension of the clamp in reference to the free thermalized stress. We do see, however,  a discrepancy in the location of this extension along the $x$-axis. Theory predicts this should be localized near the clamped side ($x=0$). In the simulated stress we instead have high extension near the edge opposite to the clamp ($x=20a$) with a region of low extension near the middle of the clamped edge. {A possible explanation of this is the auxetic behavior of a free thermalized sheet. It is well known that free thermalized polymerized sheets are controlled by a Foppl-von Karman fixed point with a negative Poison ratio.  Using a negative Poisson ratio for the stiffness matrix used to calculate our simulated stresses leads to the extensive behavior not found in typical solids. If we assume clamping renders the Poisson ratio positive (at least in a neighborhood of the clamp which can be quite large) we can recover the compressive behavior in $\sigma_{xx}$.}

\bibliographystyle{apsrev4-1}
\bibliography{refs}

\end{document}